\documentclass[11pt,twoside]{article}

\usepackage{epsfig,amsfonts,color}
\usepackage[final]{pdfpages}
\usepackage{amsmath}
\usepackage{blkarray}
\usepackage{bm}
\usepackage{algorithm}
\usepackage{algpseudocode}
\usepackage{multirow}
\renewcommand{\algorithmicfunction}{\textbf{Function}}
\algdef{SE}[TASK]{Task}{EndTask}%
   [2]{\algorithmicfunction\ \tt{#1}\ifthenelse{\equal{#2}{}}{}{(#2)}}%
   {\algorithmicend\ \algorithmicfunction}%
\usepackage{graphicx}
\usepackage[margin = 1cm]{caption}
\usepackage{subcaption}
\usepackage{placeins}

\usepackage{amssymb, palatino, geometry,url}
\usepackage[colorlinks=true,linkcolor=blue,citecolor=blue,urlcolor=blue]{hyperref}
\usepackage[titletoc,title]{appendix}

\usepackage{multirow}
\usepackage{mathtools}
\usepackage{fancyhdr}
\newcommand{\be}{\begin{equation}}
\newcommand{\ee}{\end{equation}}
\newcommand{\bea}{\begin{eqnarray}}
\newcommand{\eea}{\end{eqnarray}}

\newcommand{\bvec}{\left(\begin{array}{c}}
\newcommand{\evec}{\end{array}\right)}
\newcommand{\bsub}{\begin{subequations}}
\newcommand{\esub}{\end{subequations}}

\newcommand{\mn}{\mathcal{N}}
\newcommand{\me}{\mathcal{E}}
\newcommand{\mg}{\mathcal{G}}

\newcommand{\mrg}{\mathcal{RG}}
\newcommand{\msg}{\mathcal{SG}}

\newcommand{\bx}{\boldsymbol{x}}

\usepackage{accsupp}

\usepackage{listings}
\DeclareCaptionFont{white}{\color{white}}
\DeclareCaptionFormat{listing}{{\parbox{\dimexpr\textwidth-2\fboxsep\relax}{#1#2#3}}}
\usepackage[dvipsnames]{xcolor}

\lstdefinelanguage{julia}
{
alsoletter={!},
keywordsprefix=\@,
morekeywords={
exit,whos,edit,load,is,isa,isequal,typeof,tuple,ntuple,uid,hash,finalizer,convert,promote,
subtype,typemin,typemax,realmin,realmax,sizeof,eps,promote_type,method_exists,applicable,
invoke,dlopen,dlsym,system,error,throw,assert,new,Inf,Nan,pi,im,begin,while,for,in,return,
break,continue,macro,quote,let,if,elseif,else,try,catch,end,bitstype,ccall,do,using,module,
import,export,importall,baremodule,immutable,local,global,const,Bool,Int,Int8,Int16,Int32,
Int64,Uint,Uint8,Uint16,Uint32,Uint64,Float32,Float64,Complex64,Complex128,Any,Nothing,None,
function,type,typealias,abstract,get_node,add_edge,create_estimation_model,set_solution,
solve, get_solution, solve_ss_problem, create_estimation_problem, addnode, Partition, 
assemble_optigraph, local_subgraphs, apply_partition, OptiGraph, optimizer_with_attributes,
BendersAlgorithm, optimize, aggregate, source_graph, aggregate_to_depth, add_subgraph, fill, set_to_node_objectives, set_optimizer, RemoteOptiGraph, run_algorithm!, setdiff, all_variables, Symbol, fetch, load_operation_optinode_remote!, name, local_graph, GenericAffExpr, operation_model!, replace, intersect, findfirst, Dict
},
morekeywords = [2]{triggered_by,compute_time,trigger_during_busy,send_on,delay,send_wait,start},
sensitive=true,
morecomment=[l]{\#},
morestring=[b]',
morestring=[b]"
}

\usepackage{authblk}
\usepackage{amsthm}
\theoremstyle{plain}

\theoremstyle{definition}

\begin{document}

\title{A Graph-Based, Distributed Memory, Modeling Abstraction for Optimization}

\author{David L. Cole$^{1}$, Jordan Jalving$^{2}$, Jonah Langlieb$^{3}$, Jesse D. Jenkins$^{1,4}$\\
 {\small $^{1}$ Andlinger Center for Energy and Environment}\\
 {\small Princeton University, Princeton, NJ 08540, USA}\\
 {\small $^{2}$ Artificial Intelligence and Modeling Simulation}\\
 {\small Atomic Machines Inc., Emeryville, CA 94608, USA}\\
 {\small $^{3}$Department of Computer Science}\\
 {\small Princeton University, Princeton, NJ 08540, USA}\\
 {\small $^{4}$ Department of Mechanical and Aerospace Engineering}\\
 {\small Princeton University, Princeton, NJ 08540, USA}\\
 {\small $^*$ Corresponding author email address: jessejenkins@princeton.edu}
}

\date{}

\maketitle

\abstract{We present a general, flexible modeling abstraction for building and working with distributed optimization problems called a RemoteOptiGraph. This abstraction extends the OptiGraph model in Plasmo.jl, where optimization problems are represented as hypergraphs with nodes that define modular subproblems (variables, constraints, and objectives) and edges that encode algebraic linking constraints between nodes. The RemoteOptiGraph allows OptiGraphs to be utilized in distributed memory environments through InterWorkerEdges, which manage linking constraints that span workers. This abstraction offers a unified approach for modeling optimization problems on distributed memory systems (avoiding bespoke modeling approaches), and provides a basis for developing general-purpose meta-algorithms that can exploit distributed memory structure such as Benders or Lagrangian decompositions. We implement this abstraction in the open-source package, Plasmo.jl and we illustrate how it can be used by solving a mixed integer capacity expansion model for the western United States containing over 12 million variables and constraints.  The RemoteOptiGraph abstraction together with Benders decomposition performs 7.5 times faster than solving the same problem without decomposition.}


\section{Introduction}

Frontiers in mathematical optimization continue to shift toward ever greater scale and complexity. Practitioners often consider problems with millions (or even billions) of variables driven by demands of applications that span large-scale energy systems \cite{
hadidi2025large, zhang2024stabilised}, enterprise/area-wide scheduling and operations \cite{biegler2024multi,chouksey2025accelerated,kalamaris2025general,liu2022distributed
}, or expansive supply chains \cite{garcia2015supply,neiro2025nationwide,silva2009distributed}. 

Advances in high-performance and cloud computing (HPC and CC) \cite{kratzke2018brief,surbiryala2019cloud} have made it possible to solve increasingly larger-scale optimization problems using distributed memory and parallel computation. 
However, fully exploiting HPC and CC resources requires structure-exploiting optimization algorithms implemented in a distributed manner, which is difficult because it requires integrating optimization modeling tools with distributed memory architectures. To the authors' knowledge, no generalized distributed modeling framework for implementing such algorithms has yet emerged.

Most distributed optimization algorithms today are implemented in an ad-hoc manner, where decomposition strategies are applied on top of existing algebraic modeling languages such as JuMP \cite{Lubin2023}, Pyomo \cite{bynum2021pyomo}, AMPL \cite{fourer2003ampl}, or GAMS \cite{gams}. In these approaches, the modeling framework itself does not represent distributed structure; instead, developers implement bespoke strategies to partition models, coordinate subproblem solves, and exchange information between them, often using various tools for distributed or parallel processes (e.g., the Message Passing Interface (MPI) \cite{gropp1999using}, Dask in Python \cite{dask2016}, or Distributed.jl in Julia \cite{bezanson2017julia}). This separation creates a gap: while distributed algorithms exist, the models themselves cannot be authored natively in a distributed form. Consequently, meta-algorithms---such as Benders decomposition \cite{benders1962partitioning,rahmaniani2017benders}, alternating direction methods of multipliers (ADMM) \cite{boyd2011distributed}, or progressive hedging \cite{rockafellar1991scenarios}---are often implemented in ways that are specific to the application's problem structure and chosen modeling language. These bespoke implementations increase programming effort (e.g., repeated implementation of the same meta-algorithms in different contexts), slows the accretion of cumulative improvements in algorithmic performance and limits the broader use of successful algorithms across different contexts. These challenges can be addressed by creating a generalized abstraction for working with distributed optimization problems. 

The need for a general abstraction for distributed optimization is highlighted by several use cases that arise in literature. First, many problems have an inherent distributed communication structure \cite{bertsekas2015parallel, zheng2022review}. For example, in many power system and control problems, there is a central planner that is optimizing a larger system with limited knowledge of data on distributed entities within the system \cite{molzahn2017survey, nedic2018distributed, 
yang2019survey}. In these cases, subproblems, representing distributed entities, can be solved separately (and sometimes asynchronously) and communicate their results back to the central planner or to other subproblems \cite{assran2020advances, bertsekas2015parallel}. None of these problems, to our knowledge, use a generalized framework for modeling the distributed structure. Second, many structure-exploiting meta-algorithms \cite{
constante2025relaxation, hadidi2025large} are necessary for solving distributed problems, but these are typically implemented in a bespoke manner for specific applications. As examples, 
Benders decomposition \cite{benders1962partitioning,jacobson2024computationally,rahmaniani2017benders}, 
Lagrangian relaxations \cite{boyd2011distributed,chang2016asynchronous,
fisher1981lagrangian,wang2019distributed}, 
progressive hedging \cite{
rockafellar1991scenarios, ryan2013toward}, 
Schur decomposition \cite{cole2022julia,kang2014interior,laird2008large,word2014efficient,
zavala2008interior}, 
and Schwarz decomposition \cite{jalving2022graph,na2022convergence,
shin2020decentralized} have all been implemented in parallel fashions, but in problem-specific ways that hinder general reuse across domains. Generalizing these algorithms within a unified distributed abstraction would simplify their application to new problem domains. Third, some problems are intractable due to problem size (e.g., there can be practical memory limitations such that the solver cannot run all necessary steps to reach optimality; see for example \cite{jacobson2024computationally,zhang2024stabilised}). Solving these then can require decomposing the problem into subproblems that are run on multiple processors, requiring practitioners to integrate distributed memory tools with their algebraic modeling languages -- a specialized technical task which can be abstracted over by using a more general modeling framework.

A prominent modeling paradigm for providing flexible and general modeling capabilities is to represent optimization problems using graph theoretic concepts where optimization problems are represented by collections of nodes and edges. Multiple approaches have been proposed for capturing problem structure via graphs. Some approaches, proposed by Daoutidis and co-workers \cite{allman2019decode,daoutidis2019decomposition,mitrai2021efficient,mitrai2024computationally,tang2023resolving,tang2018optimal}, include a variable graph (where each variable is a node and edges connect variable nodes that arise in the same constraint), a constraint graph (where each constraint is a node with edges that connect nodes with a common variable), and a bipartite graph (where variables and constraints are different types of nodes). These graph representations allow for different problem decomposition schemes into subproblems via graph partitioning or community detection algorithms. The resulting subproblems can be used in various decomposition-based solution strategies including Benders decomposition. An alternative approach is presented by Jalving and co-workers \cite{jalving2022graph} which is based on an abstraction called an OptiGraph, in which nodes contain objectives, variables, and constraints, and hyperedges encapsulate constraints over variables on different nodes. This latter approach is also similar to that used by \cite{berger2021gboml}
. The OptiGraph abstraction, implemented in the open-source package Plasmo.jl, provides significant modeling flexibility as it allows for capturing hierarchical structures, supports restructuring the graph, and supports the aforementioned decomposition capabilities \cite{cole2025graph}.

In this work, we extend the OptiGraph abstraction for the case of distributed memory applications and introduce a new abstraction called a RemoteOptiGraph. The RemoteOptiGraph is implemented in the open-source tool Plasmo.jl \cite{jalving2022graph} and provides a user-friendly interface for constructing and working with distributed optimization problems. The RemoteOptiGraph abstraction closes the gap caused by the inability of models to be authored natively in a distributed form by allowing users to construct distributed optimization models with the same interface as standard OptiGraphs. Because the abstraction preserves graph-based structure, generic meta-algorithms (e.g., Benders decomposition) that operate on OptiGraphs can be applied without modification to RemoteOptiGraphs. In this way, RemoteOptiGraphs unify centralized and distributed optimization modeling, enabling algorithm developers to design reusable, graph-based methods that extend seamlessly to distributed computation, allowing practical scaling of memory and computational resources across multiple processors.

A related effort of note is MPAX \cite{lu2024mpax}, a JAX-based optimization framework that provides differentiable, hardware-accelerated solvers that are well suited for integration with modern differentiable programming workflows. MPAX distributes computation by sharding data structures across multiple devices (CPUs, GPUs, or TPUs). In this approach, large constraint matrices are partitioned across devices and collective communication is handled automatically by the underlying JAX runtime during solver execution. MPAX currently focuses on scaling numerical linear algebra for first-order methods, and specifically it provides primal–dual hybrid gradient algorithms where such sharding is most effective. RemoteOptiGraphs, on the contrary, are solver-agnostic: they leverage the broad ecosystem of optimization solvers (many accessible through JuMP.jl) and focus on distributed modeling rather than array-level sharding. The two approaches address different layers of the distributed optimization stack. MPAX advances distributed solver implementations through sharded linear algebra, whereas RemoteOptiGraphs enable distributed problem representations and coordination across subproblems, giving users explicit control over problem structure. Sharding approaches are well-suited for large continuous problems dominated by linear algebra operations, while a distributed modeling abstraction addresses the challenges of coordinating heterogeneous subproblems, managing distributed memory, and developing reusable meta-algorithms. The two frameworks are also computationally compatible and could be layered —for example, using a sharded solver to accelerate large continuous subproblems within a RemoteOptiGraph-based decomposition.

The contributions of this work include: i) we define a graph theoretic, mathematical abstraction for working with distributed optimization problems called the RemoteOptiGraph; ii) we implement this abstraction in the open-source tool Plasmo.jl, which offers a user-friendly interface for constructing distributed optimization problems while minimizing direct interaction with distributed commands and simplifying the modeling process; iii) we present a large-scale case study of a capacity expansion problem in the power system sector, and show that it can be solved with the RemoteOptiGraph and Benders decomposition more than seven times faster than solving the monolithic implementation. Our contributions ultimately highlight the ability of the RemoteOptiGraph to work with distributed problems in a seemless and intuitive fashion. While this manuscript demonstrates the development of a generic distributed Benders decomposition algorithm, the abstraction can be further applied for a variety of problems in future work. For instance, it could be used to replace parts of an optimization problem with machine-learning surrogates (see, for example, \cite{brahmbhatt2025benders}) to naturally facilitate using distributed GPU-accelerated techniques.

This paper is structured as follows. Section \ref{sec:math} introduces the mathematical abstraction of the RemoteOptiGraph. Section \ref{sec:software} describes the software implementation of the RemoteOptiGraph in Plasmo.jl and provides a short tutorial. Section \ref{sec:case_study} gives a case study for a large-scale, mixed-integer, capacity expansion problem modeled with the RemoteOptiGraph and solved with Benders decomposition. Section \ref{sec:conclusion} gives conclusions and future work.  

\section{Mathematical Abstraction}\label{sec:math}

The OptiGraph was introduced by \cite{jalving2022graph} and consists of a set of OptiNodes, $\mn$, (which can contain local variables, constraints, objective functions, and data) and a set of OptiEdges, $\me$ (which contain linking constraints for variables stored on two or more nodes). Because linking constraints may contain variables on more than two nodes, the set $\me$ contains undirected hyperedges. The OptiGraph is denoted by $\mg(\mn, \me)$. The set of OptiNodes on OptiGraph $\mg$ is denoted by $\mn(\mg)$ and the set of OptiEdges by $\me(\mg)$. The set of nodes connected by edge $e$ are given by $\mn(e)$ and the set of OptiEdges incident to OptiNode $n$ are given by $\me(n)$. When clear from context, we will omit the prefix {\it Opti} before OptiGraphs, OptiNodes, and OptiEdges. 

\begin{figure}
    \centering
    \includegraphics[width=0.8\linewidth]{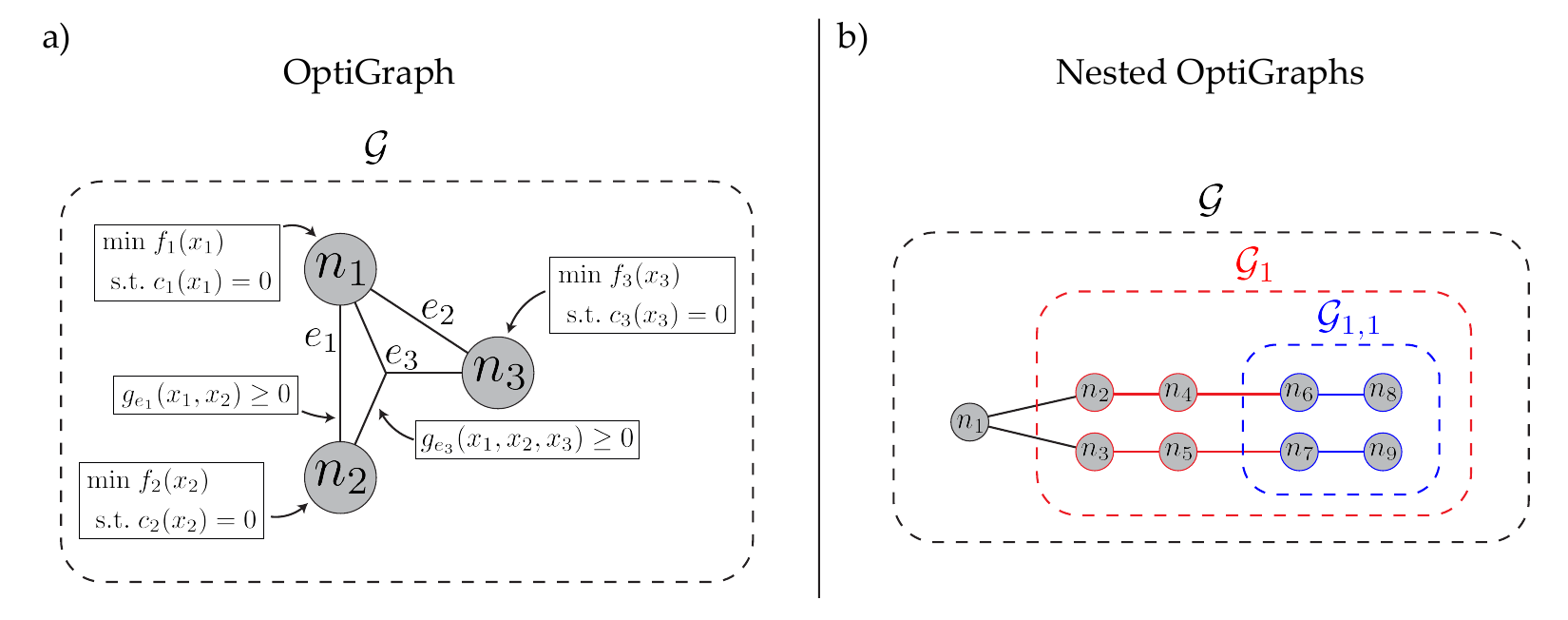}
    \caption{A visualization of the OptiGraph. Figure a shows the general abstraction with variable sets $x_1$, $x_2$, and $x_3$ partitioned into nodes $n_1$, $n_2$, and $n_3$, respectively, with edges containing linking constraints between these variable sets. Figure b shows an example of the nested structures that can be supported within OptiGraphs.}
    \label{fig:optigraph}
\end{figure}

The optimization model for an OptiGraph $\mg$ is given by: 
\begin{subequations}\label{eq:optigraph}
    \begin{align}
        \min_{\{\bx_n\}_{n \in \mn(\mg)}} &\; f\left(\{\bx_n \}_{n \in \mn(\mg)}\right) & (\textrm{Objective}) \label{eq:optigraph_objective} \\
        \textrm{s.t.} &\; \bx_n \in \mathcal{X}_n, \quad n \in \mn(\mg), \quad & (\textrm{Local Node Constraints}) \label{eq:optigraph_nodes} \\
        &\; g_e(\{\bx_n\}_{n \in \mn(e)})\geq 0, \quad e \in \me(\mg). \quad & (\textrm{Link Constraints}) \label{eq:optigraph_edges}
    \end{align}
\end{subequations}
\noindent where $\bx_n$ are the decision variables on node $n$. The objective function \eqref{eq:optigraph_objective} is a scalar composite function of any variables on nodes in the graph. \eqref{eq:optigraph_nodes} is the set of local constraints stored on nodes, and \eqref{eq:optigraph_edges} is the set of linking constraints, stored on edges. The formulation in \eqref{eq:optigraph} supports both continuous and discrete decision variables and linear and nonlinear objective functions and constraints. While \eqref{eq:optigraph_edges} are given as inequality constraints, the OptiGraph abstraction is not limited to inequality constraints as equality constraints can be rewritten as two inequality constraints. A visualization of an OptiGraph with three nodes is given in Figure \ref{fig:optigraph}a, where each node $n_1$, $n_2$, and $n_3$ contains a set of variables $x_1$, $x_2$, and $x_3$, respectively, and where each edge contains linking constraints between these variables.

Importantly, the OptiGraph abstraction also supports nesting subgraphs within other graphs, forming hierarchical structures (Figure \ref{fig:optigraph}b). In \cite{cole2025graph}, they introduced notation for these hierarchical structures, given by
\begin{equation}\label{eq:subgraph_notation}
    \mg_{\circ}(\{\mg_{\circ, i}\}_{i \in \{1, ..., N_{\circ} \}}, \mn_{\mg_{\circ}}, \me_{\mg_{\circ}}),
\end{equation}
\noindent where $\circ$ represents a set of indices which map the graph to its hierarchical level (e.g., $``$1'' for the first layer, $``$1,1'' for the first subgraph on the first layer), $N_\circ$ is the number of subgraphs on $\mg_\circ$, $\{\mg_{\circ, i}\}_{i \in \{ 1,...,N_\circ\}}$ is the set of subgraphs on $\mg_\circ$, and $\mn_{\mg_\circ}$ and $\me_{\mg_\circ}$ are the set of nodes and edges contained in $\mg_\circ$ which are not shared by their subgraphs. Under this notation, the $``$level'' of a subgraph in a hierarchical problem is given by the set of indices $\circ$. For instance, in Figure \ref{fig:optigraph}b, ${\color{red} \mathcal{G}_1}$ is nested within $\mathcal{G}$ and contains 4 nodes, while ${\color{blue} \mathcal{G}_{1,1}}$ is nested within ${\color{red} \mathcal{G}_1}$ and likewise contains four nodes. Here, the indices of these nested subgraphs indicates the subgraphs $``$level'' in the overall graph. For instance, ${\color{blue} \mathcal{G}_{1,1}}$ is notated with two indices and is two $``$levels'' deep in the graph.

\subsection{RemoteOptiGraph Abstraction}
The OptiGraph abstraction effectively captures structural relationships within an optimization problem, but it is agnostic to how those structures are realized in hardware. The abstraction assumes a shared-memory environment and provides no mechanism for managing where subgraphs or variables reside in distributed memory. While one could conceptually place subgraphs on separate workers, the original OptiGraph does not account for memory locality, communication, or data ownership across workers. Such factors become critical in distributed computing environments. As a motivating example, if two subgraphs are placed on separate workers, where do the variables of those subgraphs exist in memory? How can a linking constraint between the two subgraphs properly reference variables if those variables are stored on separate workers? The OptiGraph abstraction alone is insufficient for answering these questions and fundamentally lacks the constructs to represent or coordinate problem data residing on separate memory spaces.

To overcome this limitation, we extend the OptiGraph abstraction to operate seamlessly across distributed memory environments through the RemoteOptiGraph (which we will refer to as a ``remote graph'' where it is obvious from context). The RemoteOptiGraph allows OptiGraphs and their subgraphs to be instantiated on different workers while remaining part of a single unified optimization model. Each RemoteOptiGraph identifies its associated worker, holds a lightweight link to the OptiGraph residing on that worker, and stores only the minimal information needed for coordination, including references to sub-RemoteOptiGraphs and InterWorkerEdges (OptiEdges that connect across workers). All OptiNodes and OptiEdges remain stored on the underlying OptiGraph itself. This design preserves the simplicity of the OptiGraph interface while allowing optimization models to be constructed and solved across distributed computing environments.

We denote the RemoteOptiGraph as 
\begin{equation}\label{eq:remotesubgraph_notation}
    \mrg_{\circ}(\mg_\circ, \{\mrg_{\circ, i}\}_{i \in \{1, ..., N^{\mrg}_{\circ} \}}, \me^{IW}_\circ, w),
\end{equation}
where $\mg_\circ$ is an OptiGraph stored on remote worker $w$, $N^{\mrg}_\circ$ is the number of subgraphs contained on $\mrg_\circ$, $\{\mrg_{\circ, i}\}_{i \in \{1, ..., N^{\mrg}_{\circ} \}}$ is the set of remote subgraphs on $\mrg_\circ$,  and $\me^{IW}_\circ$ is the set of InterWorkerEdges that link between OptiGraphs stored on remote workers. Here, InterWorkerEdges--- indexed by the superscript $IW$---are a separate kind of edge from the typical OptiEdge because they can connect OptiNodes that are stored on separate RemoteOptiGraphs and therefore potentially on separate workers. In addition, each remote subgraph, $\mrg_{\circ, i}, {i \in \{1, ..., N^{\mrg}_{\circ} \}}$, has a similar form to \eqref{eq:subgraph_notation}. However, as with OptiGraphs, we denote RemoteOptiGraphs as $\mrg_\circ$ (i.e., without the parenthetical information) for simplicity. We also denote the set of subgraphs on $\mrg_\circ$ as 
\begin{equation}\label{eq:subgraph_func}
 \msg(\mrg_{\circ}) := \{\mrg_{\circ, i}\}_{i \in \{1, ..., N^\mrg_\circ \}}.
\end{equation}
\noindent Because RemoteOptiGraphs do not directly store OptiNodes or OptiEdges, we introduce the notation of $\mn(\mrg)$ and $\me(\mrg)$ as the set of OptiNodes and OptiEdges stored on the OptiGraph of the RemoteOptiGraph $\mrg$. The set of InterWorkerEdges on a RemoteOptiGraph (and its remote subgraphs if applicable) is given by $\me^{IW}(\mrg)$.

\begin{figure}
    \centering
    \includegraphics[width=0.8\linewidth]{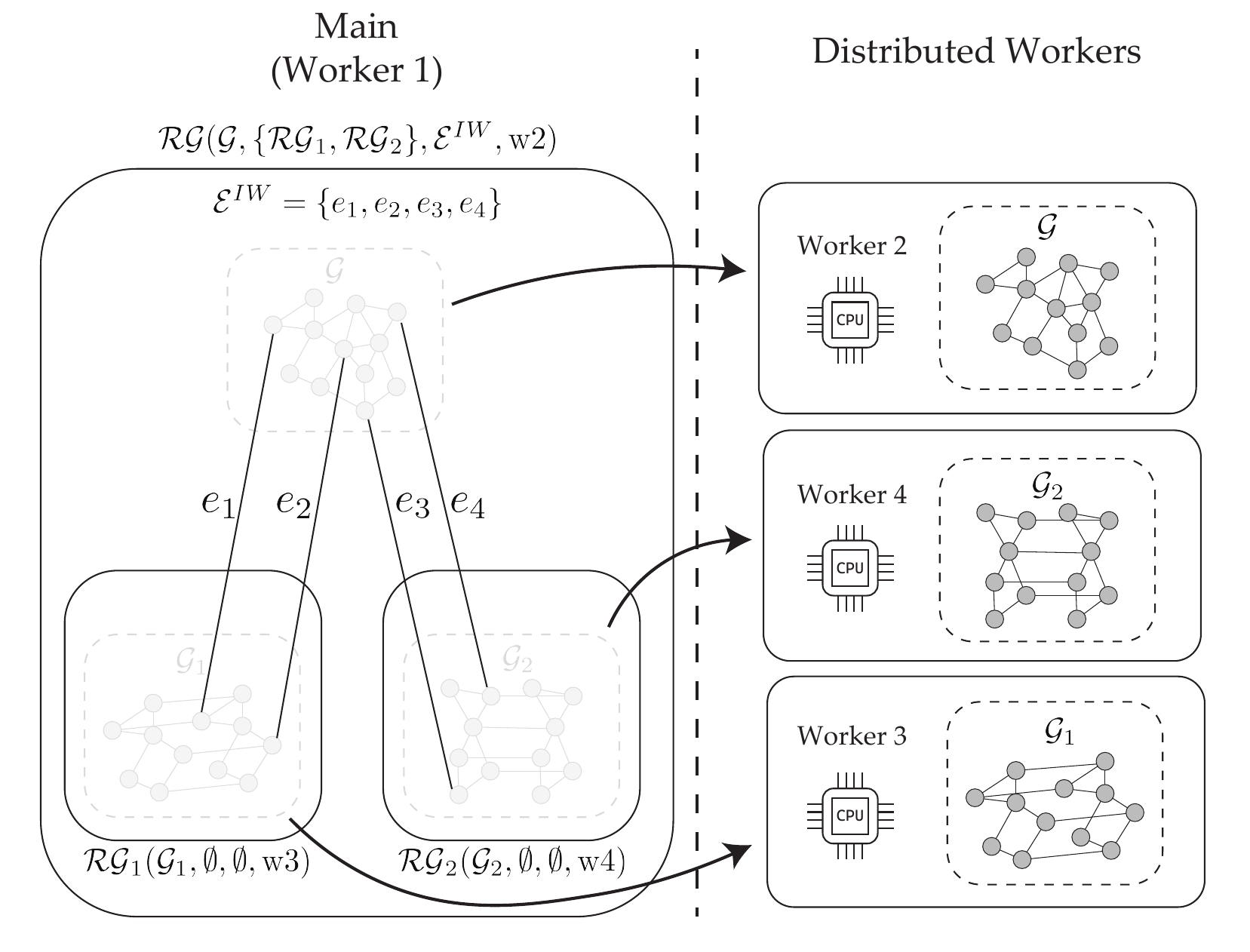}
    \caption{An example of the RemoteOptiGraph abstraction, where an overall RemoteOptiGraph, $\mrg(\mg, \{ \mrg_1, \mrg_2\}, \me_\mrg, \textrm{w2})$ is stored on the main processor with its OptiGraph, $\mg$ stored on worker 2. Two remote subgraphs, $\mrg_1$ and $\mrg_2$ also have OptiGraphs, $\mg_1$ and $\mg_2$, stored on workers 3 and 4, respectively. Only $\mrg$ contains a set of InterWorkerEdges, $\me_\mrg$.}
    \label{fig:RemoteOptiGraph}
\end{figure}

An important aspect of the RemoteOptiGraph is the location of information on different workers. The RemoteOptiGraph structure, $\mrg_\circ$, along with its InterWorkerEdges, $\me^{IW}_\circ$, are stored on the main worker, while the OptiGraph, $\mg_\circ$ is stored on some worker $w$ (this could be the main worker or a distributed worker). Each remote subgraph, $\mrg_{\circ, i}, {i \in \{1, ..., N^{\mrg}_{\circ} \}}$, likewise contains an OptiGraph that is stored on a worker (potentially different from $w$). An example of the RemoteOptiGraph abstraction is given in Figure \ref{fig:RemoteOptiGraph}. Here, there is an overall RemoteOptiGraph, $\mrg$, with an OptiGraph stored on worker 2 and with two remote subgraphs, $\mrg_1$ and $\mrg_2$, each with OptiGraphs stored on worker 3 and worker 4, respectively. Only $\mrg$ contains InterWorkerEdges, $\me^{IW}_\mrg$.

The optimization problem stored on the RemoteOptiGraph has a similar form to \eqref{eq:optigraph}. For a RemoteOptiGraph, $\mrg(\mg, \{\mrg_{ i}\}_{i \in \{1, ..., N^{\mrg} \}}, \me^{IW}, w)$, one can write the optimization formulation as 
\begin{subequations}\label{eq:remoteoptigraph}
    \begin{align}
        \min &\; f\left(\{\bx_n \}_{n \in \mn(\mg)}\right) + \sum_{g \in \msg(\mrg)} f_g\left(\{\bx_n \}_{n \in \mn(g)}\right) & (\textrm{Objective}) \label{eq:remoteoptigraph_objective} \\
        \textrm{s.t.} &\; \bx_n \in \mathcal{X}_n, \quad n \in \left(\bigcup_{g \in \msg(\mrg)} \mn(g) \right) \cup \mn(\mg) & (\textrm{Node Constraints}) \label{eq:remoteoptigraph_nodes} \\
        &\; g_e(\{\bx_n\}_{n \in \mn(e)})\geq 0, \quad e \in\left(\bigcup_{g \in \msg(\mrg)} \me(g) \right) \cup \me(\mg) & (\textrm{Local Link Constraints}) \label{eq:remoteoptigraph_edges}\\
        &\; g_e(\{\bx_n\}_{n \in \mn(e)})\geq 0, \quad e \in \me^{IW}(\mrg) & (\textrm{Inter-Worker  Link Constraints}). \label{eq:remoteoptigraph_interworker_links}
    \end{align}
\end{subequations}
\noindent In this formulation, the objective and constraints are a collection of each remote subgraph's objectives and constraints. The overall objective is a sum of the OptiGraph objectives that are stored on the original RemoteOptiGraph, $\mrg$, and the remote subgraphs. In other words, the overall objective is {\it separable} among the primary RemoteOptiGraph and its remote subgraphs, though future work could include the ability to support an InterWorkerObjective for graphs. Similarly, the constraints \eqref{eq:remoteoptigraph_nodes} and \eqref{eq:remoteoptigraph_edges} are a collection of the constraints stored on each OptiGraph of each RemoteOptiGraph. The only constraints that link between multiple OptiGraphs stored on the RemoteOptiGraphs are the constraints, \eqref{eq:remoteoptigraph_interworker_links}, which are stored on the InterWorkerEdges. These constraints can therefore contain variables stored on multiple workers.

A primary role of the RemoteOptiGraph is to provide a coherent framework for developing distributed solution schemes. While the formulation in \eqref{eq:remoteoptigraph} specifically targets the distributed setting, the abstraction itself is consistent with the original OptiGraph in the sense that it provides the same interface for building and working with optimization problems. It is straightforward for instance, to use the same model-building methods to create either an OptiGraph or RemoteOptiGraph. This common interface facilitates benchmarking distributed solutions against traditional solvers while avoiding the need to produce duplicate code. Additionally, future work will provide features that directly convert RemoteOptiGraphs into OptiGraphs (and vice versa) as another means to help benchmark and to verify against traditional optimization solvers.

\section{Software Implementation}\label{sec:software}

\subsection{Overview}

The software implementation of the RemoteOptiGraph seeks to achieve three primary functionalities. i) It allows the RemoteOptiGraph to have a ``global'' view of aspects of the optimization problem---stored across the different workers---in a memory efficient way. ii) It allows for building optimization problems on the remote workers from the main worker using standard modeling syntax. iii) It accomplishes efficient serialization and de-serialization between the main and remote workers (i.e. it uses lightweight references to communicate model-building instructions and results between workers). These goals are facilitated by two types of new data structures in Plasmo.jl: remote references and proxy references. Remote references are meant to be used from the calling worker (usually the main worker); they correspond to graph objects (like OptiGraphs and OptiNodes) on remote workers. Proxy references are used to facilitate efficient serialization/de-serialization of data between workers in a manner that supports the underlying expression types in Plasmo.jl. Both remote references and proxy references contain minimal necessary identifying information (such as UUIDs) to work with remote objects. In this section, we outline the implementation of the RemoteOptiGraph and detail the basis for the remote and proxy reference data structures.

To differentiate the generic mathematical formulation from the software implementation, we refer to the software implementation of OptiGraphs, OptiNodes, and OptiEdges and other objects in teletype font. An {\tt OptiGraph} in Plasmo.jl is comprised of sets of {\tt OptiNodes} and {\tt OptiEdges}. Plasmo.jl builds upon JuMP.jl \cite{Lubin2023}, and extends many of its functions and macros. Aspects of the implementation are ultimately what motivate the need for proxy data structures (for instance, many child objects in the OptiGraph have parent references). For a deeper coverage of the Plasmo.jl implementation, see \cite{jalving2022graph}.

The RemoteOptiGraph is implemented in Plasmo.jl beginning in version 0.7.0 (which is a precursor version to 1.0) and follows the mathematical formulation outlined in Section \ref{sec:math}.  Plasmo.jl implements the {\tt RemoteOptiGraph} by leveraging functionality in Distributed.jl \cite{bezanson2017julia} and DistributedArrays.jl \cite{distributedarrays}. As an overview of the structure, the {\tt RemoteOptiGraph} itself is stored on the main worker and includes fields for a worker (corresponding to the index of an available processor), the {\tt OptiGraph} it references on the worker (stored in a length one {\tt DistributedArray.DArray}; this {\tt DArray} is stored on the main worker as a lightweight ``pointer'' to data on the remote worker), a vector of remote subgraphs, and a vector of {\tt InterWorkerEdge} elements.

Modeling using a {\tt RemoteOptiGraph} is enabled by the introduction of remote reference and proxy reference objects. Remote reference objects include {\tt RemoteNodeRef}, {\tt RemoteEdgeRef}, and {\tt RemoteVariableRef} which correspond to the {\tt OptiNode}, {\tt OptiEdge}, and {\tt NodeVariableRef} of an {\tt OptiGraph}. These remote reference objects are used from the main worker and provide a unified, memory-efficient view of their corresponding {\tt OptiGraph} equivalents that exist on other workers. Remote references ultimately act to provide a lightweight means to work with their worker counterparts while providing the same consistent {\tt OptiGraph} modeling interface. While the remote references are user facing, proxy reference objects are used internally (i.e., not user-facing) in Plasmo.jl to  facilitate memory-efficient serialization and de-serialization. Proxy reference objects include the {\tt ProxyNodeRef}, {\tt ProxyEdgeRef}, and {\tt ProxyVariableRef} which likewise correspond to nodes, edges, and variables of an {\tt OptiGraph}. Because {\tt RemoteOptiGraph} or {\tt OptiGraph} objects (e.g., an {\tt OptiNode}) contain a reference to their parent graph, passing these objects between workers would de-reference their entire parent  and require significant serialization, which is not practical. To avoid this serialization, each proxy reference stores only minimal identifying information (i.e., they strip the parent reference to the {\tt OptiGraph} or {\tt RemoteOptiGraph} to avoid serializing it) such as node identifiers, variable names, and variable indices, each of which act as an inexpensive handle to pass between workers.

Similar to the {\tt OptiGraph}, the {\tt RemoteOptiGraph} extends core JuMP.jl functionality. A significant feature of JuMP.jl is the user-friendly interface for constructing optimization models, and Plasmo.jl leverages JuMP.jl's modeling functionality by extending key macros such as {\tt @variable}, {\tt @constraint}, {\tt @objective}, and {\tt @expression}, as well as functions for working with models (nodes or graphs), variables, and constraints, such as {\tt objective\_value}, {\tt set\_start\_value}, or {\tt dual}. The {\tt RemoteOptiGraph} provides the same syntax as an {\tt OptiGraph} by likewise extending these functions for the distributed case using {\tt RemoteNodeRef}, {\tt RemoteEdgeRef}, {\tt RemoteVariableRef}, or their derivative expressions (e.g., an expression containing one or more {\tt RemoteVariableRef}). In other words, calling {\tt @variable} on a {\tt RemoteNodeRef} will add a variable to the {\tt OptiNode} and return a {\tt RemoteVariableRef}. To perform this task, Plasmo.jl references the worker storing the {\tt OptiNode}, adds a variable to the {\tt OptiNode}, and then returns the {\tt RemoteVariableRef} which corresponds to that new variable. This task includes sending instructions to the worker and fetching the resulting response. This is handled internally, requiring almost no direct interaction with Distributed.jl and providing a user-friendly modeling experience. Note that because using the extended JuMP.jl functions and macros for the {\tt RemoteOptiGraph} requires serialization to remote workers, repeated calls (e.g., for building very large models) can lead to notable communication overhead. In the Appendix, we discuss this in more depth and give examples of multiple ways of constructing the {\tt RemoteOptiGraph}, with and without these extended capabilities, using the case study example in Section \ref{sec:case_study}.

\begin{figure}
    \centering
    \includegraphics[width=1\linewidth]{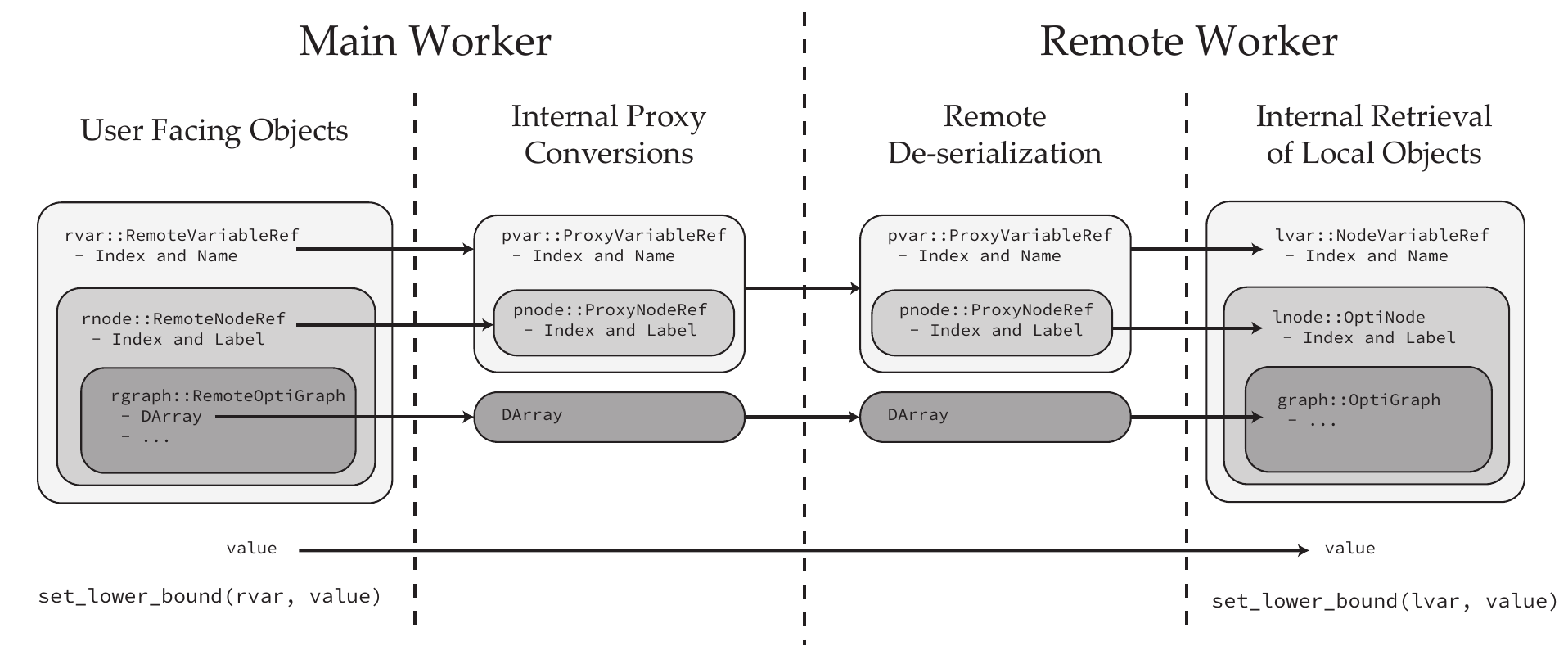}
    \caption{An example showing communication between the main and remote workers via proxy objects. When {\tt set\_lower\_bound} is called, the user-facing, remote reference objects (left) are converted internally to proxy objects, passed to the remote workers, and used to retrieve their corresponding local objects. Here the {\tt DArray} is a lightweight ``pointer'' for the {\tt OptiGraph} stored remotely.}
    \label{fig:set_lower_bound}
\end{figure}

The process of moving from the main worker to another worker is visualized in Figure \ref{fig:set_lower_bound} for the example of calling {\tt set\_lower\_bound} (a function originally extended from JuMP.jl). Here, the user calls the function with arguments  {\tt rvar} (a {\tt RemoteVariableRef}) and {\tt value} (the float value to be set as the lower bound). Internally, Plasmo.jl converts the {\tt rvar} into a {\tt ProxyVariableRef}, transfers that proxy variable to the remote worker (via serialization and de-serialization), retrieves the corresponding {\tt NodeVariableRef}, and sets the lower bound of that variable to {\tt value}. All of the proxy conversions and local variable retrieval are performed internally and are not exposed to the user. The original {\tt rvar} visualized in the Figure contains references to a {\tt RemoteNodeRef} which contains a reference to the {\tt RemoteOptiGraph}. To avoid serializing the {\tt RemoteOptiGraph}, the proxy objects are created (stripping the reference to the graph), and these are passed to the remote worker, along with the {\tt DArray} object from the original {\tt RemoteOptiGraph}. The {\tt ProxyVariableRef} contains a single index (an integer) and name (a string) while the {\tt ProxyNodeRef} contains a node index (a UUID as a symbol) and a label (also a symbol). The {\tt DArray} object is essentially a pointer to the {\tt OptiGraph} on the remote worker, so each of these three objects can be serialized and transferred to the remote worker for little expense.  

\subsection{Building a RemoteOptiGraph} \label{sec:tutorial}

\begin{figure}
    \centering
    \includegraphics[width=1\linewidth]{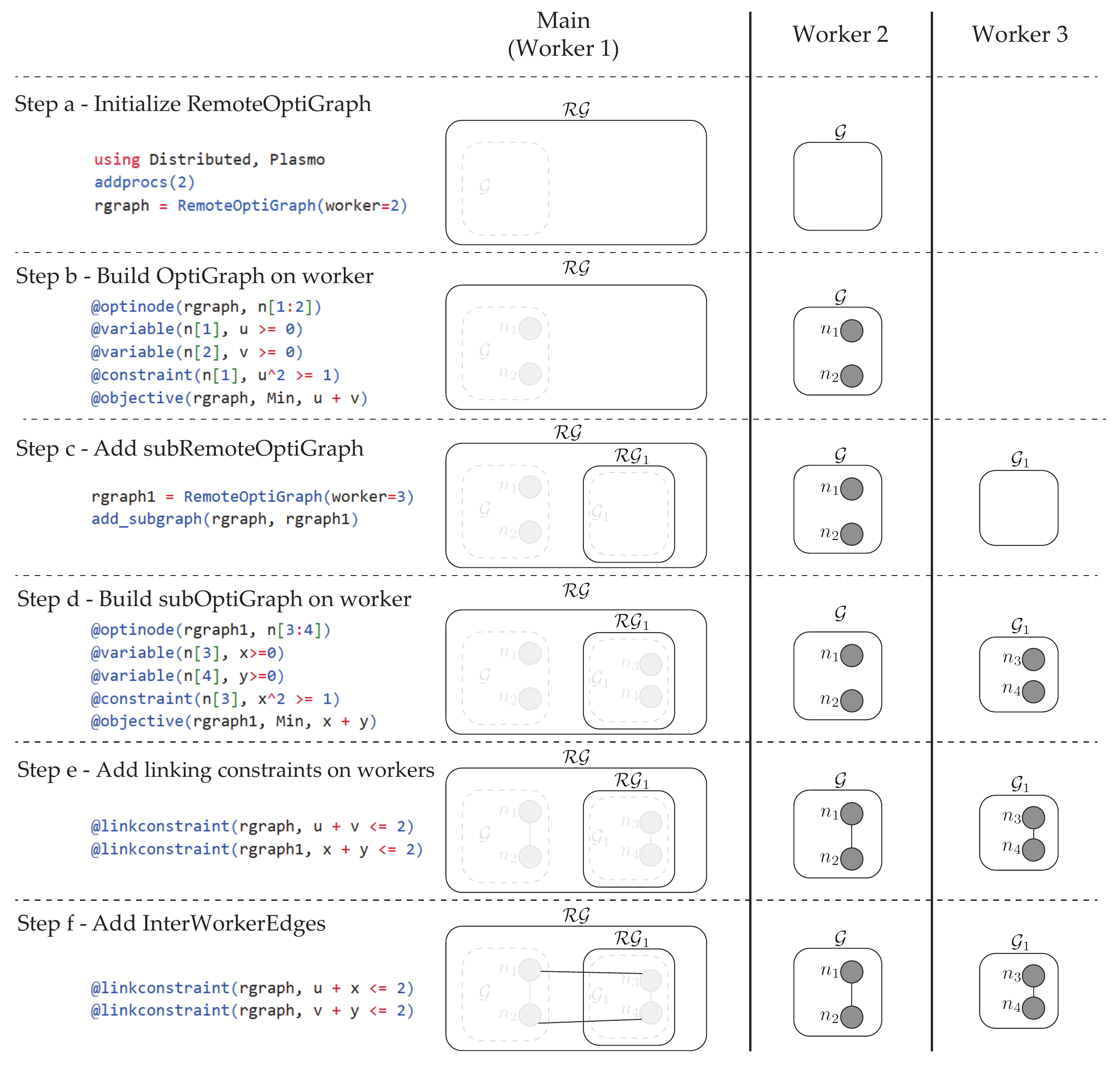}
    \caption{A visual example of building a {\tt RemoteOptiGraph} in {\tt Plamso.jl}.} 
    \label{fig:tutorial}
\end{figure}

We now give a tutorial overview of how a {\tt RemoteOptiGraph} can be constructed following the steps in Figure \ref{fig:tutorial}. To begin, Step a adds two processes (workers) to the Julia environment via {\tt addprocs} (as is typical when using the Distributed.jl package) which enables the same code to run seamlessly across any Distributed.jl supported configuration including one multi‑core machine, high‑performance clusters, or cloud computing environments, as discussed in the appendix. It then instantiates a {\tt RemoteOptiGraph} called {\tt rgraph} by calling {\tt RemoteOptiGraph{(worker=2)}}, where the {\tt worker} keyword argument denotes the process id on which the underlying {\tt OptiGraph} is stored. Step b uses the macros {\tt @optinode}, {\tt @variable}, {\tt @constraint}, and {\tt @objective} to build up the {\tt OptiGraph} on worker 2 (where we highlight the use of standard Plasmo.jl/JuMP.jl syntax). The syntax here is made possible by appropriately extending {\tt RemoteOptiGraph} methods such that they return aforementioned remote references. Step c creates an additional {\tt RemoteOptiGraph} on worker 3 called {\tt rsub} and then adds it as a subgraph to {\tt rgraph}. Step d then builds the model on {\tt rsub}. Step e adds linking constraints to both {\tt rgraph} and {\tt rsub} that connect nodes within those graphs (which each return a {\tt RemoteEdgeRef} because the constraints are local to each process). Step f adds linking constraints that connect nodes across remote graphs (which each return an {\tt InterWorkerEdge} because the constraints span across workers).

\subsection{Optimal Sizing Example}

As a more complete tutorial, we reproduce a sizing and inventory optimization problem used by \cite{cole2025graph} for a process that converts raw material $X$ into product $Y$. Because product prices fluctuate and production and sales are constrained, the model optimizes the capacity and use of storage to maximize profit over time.

The mathematical formulation is given by: 
\begin{subequations}\label{eq:storage_problem}
    \begin{align}
        \min &\; \alpha\cdot s_{size} + \sum^T_{t = 1} \left( \beta_t\cdot x^{buy}_t - \gamma_t\cdot y^{sell}_t \right) \\
        &\; y^{stored}_{t+1} - y^{stored}_t = y^{save}_t, t = 1, ..., T-1 \label{eq:storage_mass_balance}\\
        &\; y^{save}_t + y^{sell}_t - \zeta x^{buy}_t = 0, t = 1, ..., T \label{eq:y_mass_balance} \\
        &\; 0 \le y^{stored}_t \le s_{size}, t = 1, ..., T \label{eq:storage_upper_bound} \\
        &\; 0 \le y^{sell}_t \le \overline{d}^{sell}, t = 1, ..., T \\
        &\; \underline{d}^{save} \le y^{save}_t \le \overline{d}^{save}, t = 1, ..., T \\
        &\; y^{stored}_1 = \bar{y}^{stored}
    \end{align}
\end{subequations}
\noindent where the decision variables are the maximum storage size, $s_{size}$; the amount of raw material purchased, $x^{buy}_t$; the amount of product sent to storage (can be negative for product removal from storage), $y^{save}_t$; the amount of product sold $y^{sell}_t$; and the amount of product in storage, $y^{stored}_t$. $\alpha$ is the cost of building the storage, $\beta_t$ is the cost of buying $X$, $\gamma_t$ is the cost of selling $Y$, and $\zeta$ is a conversion factor from $X$ to $Y$. The parameters $\overline{d}^{sell}$, $\underline{d}^{save}$, and $\overline{d}^{save}$ are upper and lower bounds on their respective variables, while $\overline{y}^{stored}$ is the initial amount in storage. In the formulation, constraints \eqref{eq:storage_mass_balance} and \eqref{eq:y_mass_balance} are mass balances on storage and $Y$, respectively, and constraint \eqref{eq:storage_upper_bound} limits the amount of $Y$ in storage by $s_{size}$.

The problem in \eqref{eq:storage_problem} is implemented as a {\tt RemoteOptiGraph} in Code Snippet \ref{code:plasmo_example}. This problem is hierarchical in the sense that there is a planning decision ($s_{size}$) that influences the operational decisions ($x^{buy}_t$, $y^{save}_t$, $y^{sell}_t$, and $y^{stored}_t$). In building this problem, we place each hierarchical layer of the problem on a separate {\tt RemoteOptiGraph}. That is, we place the planning variables on one {\tt RemoteOptiGraph} and the operational variables on another. To begin, we first add two additional processes (Line \ref{line:addprocs}) and then ensure necessary Julia packages are loaded on each process in the standard way (Line \ref{line:everywhere}). We then set the problem data (Lines \ref{line:data_begin}-\ref{line:data_end}) and instantiate a new overall {\tt RemoteOptiGraph} called {\tt graph} (Line \ref{line:overallgraph}). Two additional subgraphs are instantiated (Lines \ref{line:planninggraph} and \ref{line:operationgraph}), one on worker 2 for the planning problem and the other on worker 3 for the operations problem. These graphs are then added as subgraphs to {\tt graph} (Line \ref{line:add_subgraph}). The planning problem is then constructed using Plasmo.jl syntax (Lines \ref{line:planningnodes_begin}-\ref{line:planningnodes_end}), followed by the construction of the operational problem (Lines \ref{line:operationnodes_begin}-\ref{line:operationnodes_end}). Linking constraints corresponding to \eqref{eq:storage_upper_bound} are added next (Line \ref{line:link_layers}), which implicitly create {\tt InterWorkerEdges}. Once this is done, the objective function and optimizer are set on each remote subgraph (Lines \ref{line:obj_and_optimizer_begin}-\ref{line:obj_and_optimizer_end}).

\begin{figure}[!htp]
    \begin{minipage}[t]{1\linewidth}
        \begin{scriptsize}
        \lstset{language=Julia, breaklines = true}
        \begin{lstlisting}[label = code:plasmo_example, caption = Code for generating the {\tt RemoteOptiGraph} of the storage problem \eqref{eq:storage_problem} in Plasmo.jl.] 
using Plasmo, HiGHS, Distributed

addprocs(2) # Uses default multiprocess (single node) parallelism |\label{line:addprocs}|
@everywhere  using Plasmo, HiGHS, Distributed |\label{line:everywhere}|
   
# Set Problem data
T = 20 |\label{line:data_begin}|
gamma = fill(5, T); beta = fill(20, T); alpha = 10; zeta = 2
gamma[8:10] .= 20; gamma[16:20] .= 50; d_sell = 50; d_save = 20; d_buy = 15; y_bar = 10 |\label{line:data_end}|
            
# Instantiate remote graphs
graph = RemoteOptiGraph(worker = 1) |\label{line:overallgraph}|
planning_graph = RemoteOptiGraph(worker = 2) |\label{line:planninggraph}|
operation_graph = RemoteOptiGraph(worker = 3) |\label{line:operationgraph}|
add_subgraph(graph, planning_graph); add_subgraph(graph, operation_graph) |\label{line:add_subgraph}|

# Define planning node data which includes storage size
@optinode(planning_graph, planning_node) |\label{line:planningnodes_begin}|
@variable(planning_node, storage_size >= 0)
@objective(planning_node, Min, storage_size * alpha) |\label{line:planningnodes_end}|

# Define operation nodes, loop through and set variables, constraint, and objective
@optinode(operation_graph, operation_nodes[1:T]) |\label{line:operationnodes_begin}|

for (j, node) in enumerate(operation_nodes)
    @variable(node, 0 <= y_stored)
    @variable(node, 0 <= y_sell <= d_sell)
    @variable(node, -d_save <= y_save <= d_save)
    @variable(node, 0 <= x_buy <= d_buy)
    @constraint(node, y_save + y_sell - zeta * x_buy == 0)
    @objective(node, Min, x_buy * beta[j] - y_sell * gamma[j])
end

# Set initial storage level
@constraint(operation_nodes[1], operation_nodes[1][:y_stored] == y_bar)
            
# Set mass balance on storage unit
@linkconstraint(operation_graph, [i = 1:(T - 1)], operation_nodes[i + 1][:y_stored] -
                        operation_nodes[i][:y_stored] == operation_nodes[i + 1][:y_save]) |\label{line:operationnodes_end}|

# Link planning decision to operations decisiosn
@linkconstraint(graph, [i = 1:T], operation_nodes[i][:y_stored] <= planning_node[:storage_size]) |\label{line:link_layers}|
            
# Set graph objective
set_to_node_objectives(planning_graph); set_to_node_objectives(operation_graph) |\label{line:obj_and_optimizer_begin}|
set_optimizer(planning_graph, HiGHS.Optimizer)
set_optimizer(operation_graph, HiGHS.Optimizer) |\label{line:obj_and_optimizer_end}|
        \end{lstlisting}
        \end{scriptsize}
    \end{minipage}
\end{figure}

\begin{figure}
    \centering
    \includegraphics[width=0.7\linewidth]{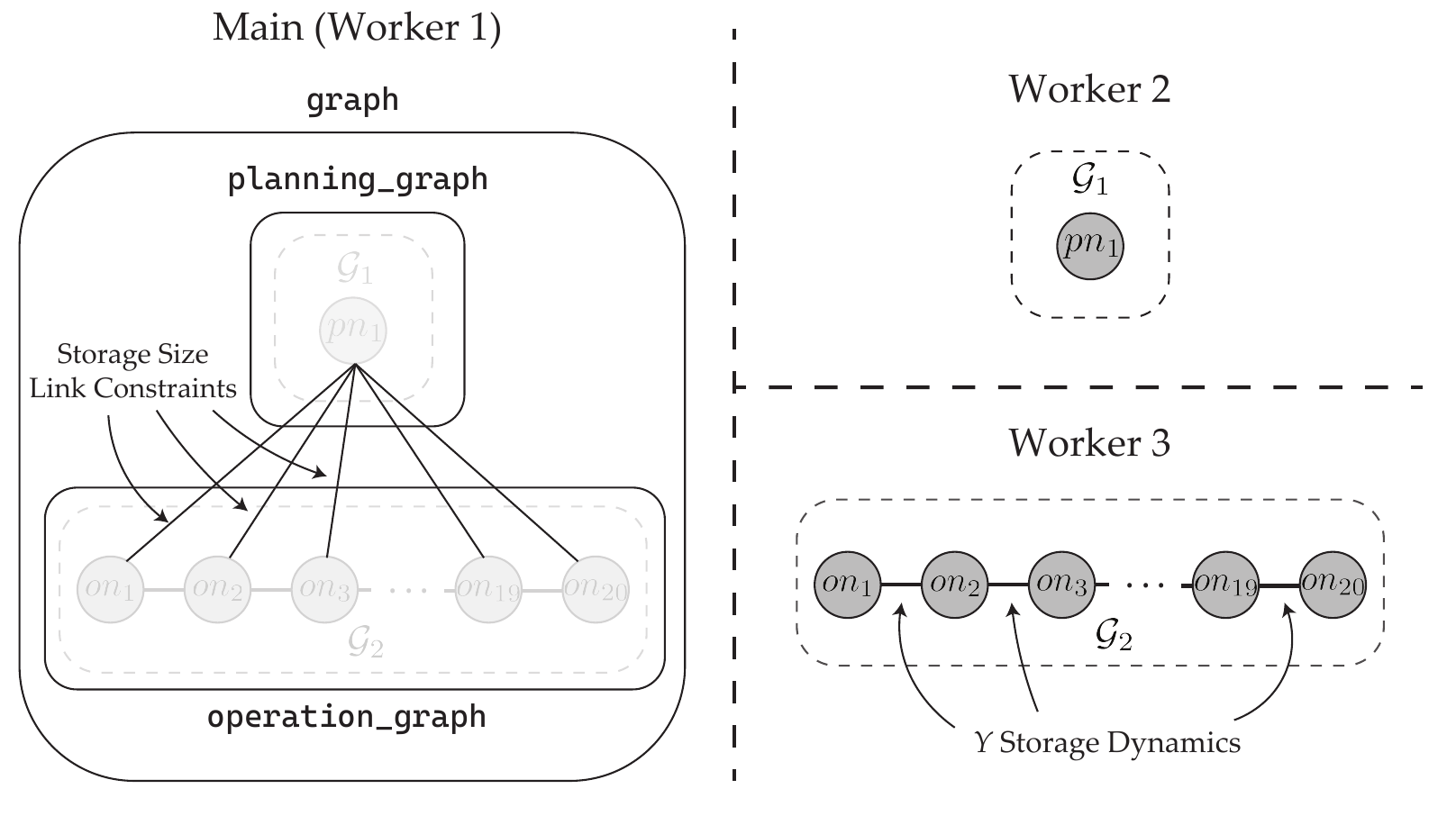}
    \caption{A visualization of the graph created by Code Snippet \ref{code:plasmo_example}, with a {\tt RemoteOptiGraph} named {\tt graph} and two remote subgraphs named {\tt planning\_graph} and {\tt operation\_graph}.}
    \label{fig:storage_example}
\end{figure}

The resulting {\tt RemoteOptiGraph} of Code Snippet \ref{code:plasmo_example} is visualized in Figure \ref{fig:storage_example}. Here, the object {\tt graph} is stored on the main worker (where its corresponding {\tt OptiGraph} appears empty as it only contains subgraphs). The two remote subgraphs {\tt planning\_graph} and {\tt operation\_graph} objects each reference an {\tt OptiGraph} stored on workers 2 and 3, respectively. The {\tt InterWorkerEdges} in {\tt graph} on the main worker contain the storage size linking constraints of \eqref{eq:storage_upper_bound}, while the storage dynamics linking constraints for \eqref{eq:storage_mass_balance} are contained in the {\tt OptiGraph} of {\tt operation\_graph} stored on worker 3. 

\subsection{Decomposition with RemoteOptiGraphs}

A {\tt RemoteOptiGraph} is intended to be used in a decomposition-based algorithm or other specialized solution approach. Because the optimization problem data is distributed across multiple processes, the overall problem is not meant to be solved with standard, shared-memory solvers as can be done with an {\tt OptiGraph} (although an {\tt OptiGraph} equivalent could easily be constructed with the same syntax for benchmarking purposes). Instead, each remote subgraph is meant to be solved as part of a coordinated scheme across workers. Here we highlight one example of how this can be done with Benders decomposition, but the principles can be applied to other decomposition or approximation strategies (e.g., ADMM).

In \cite{cole2025graph}, the authors introduced a Benders decomposition algorithm based on {\tt OptiGraphs} which was implemented in PlasmoBenders.jl. This algorithm is naturally extended to work with a {\tt RemoteOptiGraph} because it remains consistent with the OptiGraph abstraction. So long as the algorithm's structural criteria given in \cite{cole2025graph} are met (e.g., linking edges do not connect more than two subgraphs), the user can execute the algorithm on a {\tt RemoteOptiGraph} just as they would a standard {\tt OptiGraph}. An example of this approach for solving the problem in Code Snippet \ref{code:plasmo_example} with PlasmoBenders.jl is given in Code Snippet \ref{code:plasmobenders_example}. PlasmoBenders.jl supports using a {\tt RemoteOptiGraph} as of version 0.3.0.

\begin{figure}[!htp]
    \begin{minipage}[t]{0.9\linewidth}
        \begin{scriptsize}
        \lstset{language=Julia, breaklines = true}
        \begin{lstlisting}[label = code:plasmobenders_example, caption = {Code for solving example \eqref{eq:storage_problem}, constructed in Code Snippet \ref{code:plasmo_example}, using the graph-based Benders Decomposition implemented in PlasmoBenders.jl}] 
using PlasmoBenders

benders_alg = BendersAlgorithm(
    graph, # Overall graph
    planning_graph; # "root" graph
    # additional keyword arguments if desired
    add_slacks = true # add slacks to ensure recourse
)

run_algorithm!(benders_alg)
        \end{lstlisting}
        \end{scriptsize}
    \end{minipage}
\end{figure}



\section{Case Study}\label{sec:case_study}

In this section, we present a case study of an electricity system capacity expansion model (CEM), built with GenX.jl \cite{GenX}, that can be solved using Benders decomposition via PlasmoBenders.jl and the RemoteOptiGraph abstraction. This model seeks to determine electricity generation as well as storage investment and retirement decisions considering operations at an hourly resolution. Planning decisions are discrete (integer) in this case (e.g., how many power plants to build) while operational decisions are continuous (e.g., how much power to produce at each power plant). This case study considers the western area of the United States, split into 6 zones (see \cite{jacobson2024computationally}) with transmission power flow constraints between each zone and 52, week-long, operational subperiods. The optimization problem contains 12.17 million variables (360 integer) and 16.03 million constraints, was a test case solved by Pecci and co-workers \cite{pecci2025regularized}, and is efficiently solved with Benders decomposition. This case study highlights how the RemoteOptiGraph abstraction can facilitate modeling and solving large-scale, distributed optimization problems and can be used within structure-exploiting algorithms like Benders decomposition.

The form of this problem is given below, largely following the model defined by \cite{cole2025graph, pecci2025regularized}:
\begin{subequations}\label{eq:CEM}
\begin{alignat}{3}
    \min\ & c_p^\top x_p + \sum_{w \in W} c_w^\top x_w &&& \label{eq:CEM_objective}\\
    \textrm{s.t.}\ & A_w x_w + B_w x_p \le b_w, \quad && w &\ \in W \label{eq:CEM_linking}\\
    & q_w \le d, \quad && w &\ \in W \label{eq:CEM_policy1}\\
    & Q_w x_w \le q_w, \quad && w &\ \in W \label{eq:CEM_policy2}\\
    & Dx_w \le f_w, \quad && w &\ \in W \label{eq:CEM_operational_constraints}\\
    & x_w \in \mathcal{X}_w, \quad && w &\ \in W \\
    & x_p \in \mathcal{X}_p &&&
\end{alignat}
\end{subequations}


\noindent where $W$ is the set of operational subproblems (e.g., weeks); $x_p$ are the set of planning decision variables (e.g., investment and retirement decisions), including integer variables; $x_w$ are the operational decision variables of subproblem $w$; $q_w$ are a set of policy decision variables (such as allocation of CO$_2$ emissions limits); $\mathcal{X}_p$ and $\mathcal{X}_w$ are feasible regions for their respective variables; and $c_p$, $c_w$, $A_w$, $B_w$, $b_w$, $d$, and $Q_w$ are problem data. The objective \eqref{eq:CEM_objective} defines total system cost. Constraints \eqref{eq:CEM_linking} are linking constraints between the planning and operational subperiods (e.g., increased production limits from investing in new generators), constraints \eqref{eq:CEM_policy1} and \eqref{eq:CEM_policy2} are policy constraints that couple operational subperiods (e.g., CO$_2$ emission limits or renewable portfolio standards), and constraints \eqref{eq:CEM_operational_constraints} are operational constraints, such as transmission or ramping limits that are fully contained within a given subperiod. Benders decomposition decomposes this problem into a master problem (containing variables $x_p$ and $q_w$) and $|W|$ subproblems (each containing variables $x_w$). The introduction of the policy variables $q_w$ allow for eliminating linkages between subproblems (see \cite{jacobson2024computationally}). For the 6-zone case we consider here, $|W| = 52$ for the 52, week-long subperiods.

\begin{figure}
    \centering
    \includegraphics[width=0.6\linewidth]{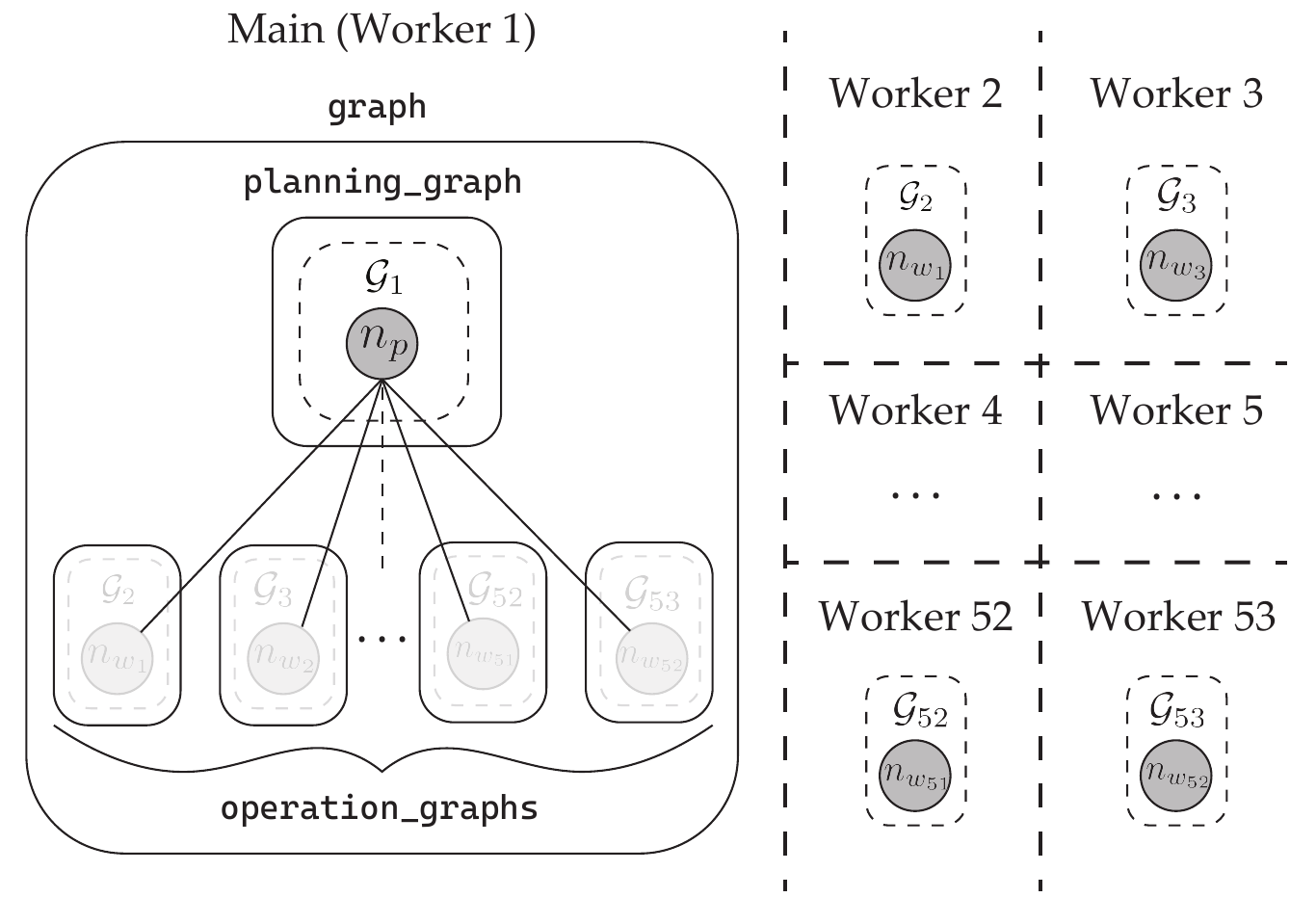}
    \caption{Structure of the 6-zone CEM built with the RemoteOptiGraph abstraction. The overall RemoteOptiGraph {\tt graph} contains a remote subgraph {\tt planning\_graph} and 52 remote subgraphs contained in the vector {\tt operation\_graphs}, each corresponding to a different worker.}
    \label{fig:CEM_structure}
\end{figure}

The 6-zone CEM, following \eqref{eq:CEM}, is modeled using the RemoteOptiGraph abstraction in Plasmo.jl and visualized in Figure \ref{fig:CEM_structure}. The overall RemoteOptiGraph, {\tt graph}, contains subgraphs for the Benders master problem and subproblems. The variables $x_p$ and $q_w, w \in W$ are placed on a planning node, $n_p$ on the remote subgraph {\tt planning\_graph}. The OptiGraph of this sub-RemoteOptiGraph is stored on the main worker. Each subperiod of the problem is treated as a subproblem and placed on their own remote subgraph in the vector {\tt operation\_graphs}. Each of these subgraphs contains a single node, $n_w$, for each $w \in W$. Each of these subgraphs is placed on a worker. We chose to solve this problem with 53 workers, such that workers 2 through 53 each contain one subproblems.
%
The assignment of subproblems to workers is arbitrary by default, and the RemoteOptiGraph abstraction remains entirely agnostic to this distribution. In this sense, the abstraction is flexible (able to accommodate a wide range of deployment configurations) and generic (independent of specific hardware or architecture). Advanced users may explicitly control worker placement when desired, but the default arbitrary assignment is sufficient for most applications.
%
InterWorkerEdges, stored on the main worker, are placed between the master problem and each subproblem and contain the linking constraints \eqref{eq:CEM_linking} and \eqref{eq:CEM_policy2}. Further details on the code for creating this RemoteOptiGraph structure is found in the appendix. Code for replicating these results and for building the tutorials can be found at \cite{Zenodo_data}.

\begin{figure}
    \centering
    \includegraphics[width=0.6\linewidth]{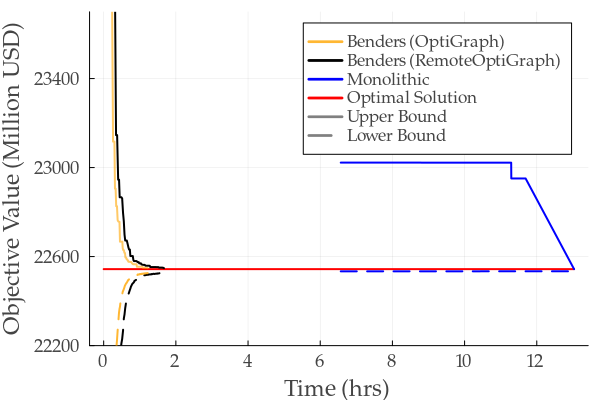}
    \caption{Results for solving the 6-zone CEM using Benders decomposition for both the RemoteOptiGraph and the OptiGraph as well as solving the OptiGraph without decomposition (``Monolithic''). Upper bounds are shown by solid lines and lower bounds are shown by dashed lines. Each problem is solved to a 0.1\% gap.}
    \label{fig:Solution_times}
\end{figure}

The above RemoteOptiGraph implementation was solved with Benders decomposition using PlasmoBenders.jl, and computational results are shown in Figure \ref{fig:Solution_times}.
For comparison, the problem was also implemented as an (non-Remote) OptiGraph stored entirely in shared memory on a single node, with an identical graph structure as the RemoteOptiGraph but where subgraphs are no longer stored on workers and InterWorkerEdges are replaced with OptiEdges. This shared‑memory implementation is practical only for such relatively small models; for large‑scale problems, it can become impractical, if not impossible, to execute as the memory requirements exceed what a single machine can reasonably provide. The OptiGraph implementation was solved using Benders decomposition in PlasmoBenders.jl and solved as a monolithic problem without Benders decomposition.
In all cases, 53 threads were used on the same compute node and solved to a 0.1\% gap on AMD EPYC 96 core, 2.4GHz processors with Gurobi version 12.0.0 \cite{gurobi} as the optimizer.
The RemoteOptiGraph implementation took 1.67 hours to solve with Benders. The (non-Remote) OptiGraph implementation took 1.22 hours to solve with Benders and 13.03 hours to solve without decomposition. enders decomposition for both the RemoteOptiGraph and the OptiGraph took 7.5 and 10.7 times less time, respectively, to solve than the monolithic problem. In fact, both reach a solution before the monolithic problem finds an incumbent solution (see Figure \ref{fig:Solution_times}).
The longer runtime of Benders observed for the RemoteOptiGraph reflects communication overhead inherent to distributed execution, even on a single machine.
This difference is expected and reflects a natural trade‑off: the RemoteOptiGraph abstraction incurs modest communication overhead in exchange for the ability to operate seamlessly across multi‑node and heterogeneous computing environments without modification to user code.

To further test the RemoteOptiGraph implementation, the RemoteOptiGraph formulation was also solved on i) separate compute nodes in the same cluster using the same processors as for the single node instance and ii) using cloud computing. These results are not included in Figure \ref{fig:Solution_times} because of the difference in computing architectures. However, both solved successfully in similar times as the results shown above. Importantly, the code for constructing and solving the RemoteOptiGraph objects on both of these architectures was identical to that used for solving on a single node. The difference in the code was how processors are added via the Distributed.jl package. These differences are discussed in the appendix and in the repository containing our code \cite{Zenodo_data}.

This case study highlights several important contributions of this work. First the RemoteOptiGraph abstraction does provide a general framework for modeling large-scale, complex optimization problems using distributed memory. In this case study, the OptiGraph implementation did solve faster with Benders. While not all problems can be solved in shared memory, we observe that the communication overhead incurred by the distributed implementation was small compared to the time saved from solving this problem with decomposition rather than without. Second, the distributed framework is agnostic to underlying architecture in the sense that the code for constructing or solving the RemoteOptiGraph is virtually identical for solving on a single compute node, multiple compute nodes, or on CC architectures. Third, this case study highlights that structure-exploiting algorithms, like Benders decomposition, can be generalized to the graph-based case, enabling the algorithm to be applied to a variety of applications so long as they are modeled as a graph with the correct structure. While this paper focused on Benders decomposition, other algorithms could be implemented, such as ADMM. Finally, this case study gave an example of a large-scale program that can be solved more than seven times faster using decomposition strategies than without, and showed that this can be done efficiently using the general RemoteOptiGraph framework. Overall, the RemoteOptiGraph abstraction provides a generalized and user-friendly framework for building distributed optimization problems and which can be used as a basis for meta-algorithms.

\section{Conclusions and Future Work}\label{sec:conclusion}

In this work, we have introduced the RemoteOptiGraph abstraction which provides a general, flexible modeling framework for distributed optimization problems. It is implemented in the open-source package, Plasmo.jl and provides a user-friendly interface for constructing optimization problems in distributed memory architectures. We have shown that it works with the open-source package, PlasmoBenders.jl, to solve distributed optimization problems using Benders decomposition, providing the user with access to decomposition schemes without implementing their own, application-specific, bespoke algorithms. Other algorithms could likewise be generalized to the RemoteOptiGraph abstraction. We have also shown that the abstraction can be used for solving large-scale optimization problems using decomposition algorithms, and we show an example of this for a CEM with over 12 million variables and 16 million constraints, which solves more than seven times faster using the RemoteOptiGraph and Benders decomposition than solving the model monolithically.

In future work, we would like to expand on the capabilities in Plamso.jl. In general, further advancements could be made within PlasmoBenders.jl, such as supporting automatically relaxing the integer variables of the master problem to generate ``warm-starting'' cuts. Other algorithms, such as ADMM, could also be implemented in a general way for the RemoteOptiGraphs. We would also like to add additional partitioning capabilities within the RemoteOptiGraph. Currently, Plasmo.jl supports partitioning an existing OptiGraph into subgraphs and then distributing those subgraphs to remote workers to form a RemoteOptiGraph, but there is not currently a way to repartition a RemoteOptiGraph across workers. Lastly, we are interested in using the RemoteOptiGraph for real-world applications, including power and energy system modeling.

\section*{Acknowledgments}
This work was supported by a grant from Clean Grid Initiative and Breakthrough Energy

\bibliography{DistributedPlasmo}

\end{document}


\sloppy

\title{Appendix for\\$``$A Graph-Based, Distributed Memory, Modeling Abstraction for Optimization''}

\author{David L. Cole$^{1}$, Jordan Jalving$^{2}$, Jonah Langlieb$^{3}$, Jesse D. Jenkins$^{1,4}$\\
 {\small $^{1}$ Andlinger Center for Energy and Environment}\\
 {\small Princeton University, Princeton, NJ 08540, USA}\\
 {\small $^{2}$ Artificial Intelligence and Modeling Simulation}\\
 {\small Atomic Machines Inc., Emeryville, CA 94608, USA}\\
 {\small $^{3}$Department of Computer Science}\\
 {\small Princeton University, Princeton, NJ 08540, USA}\\
 {\small $^{4}$ Department of Mechanical and Aerospace Engineering}\\
 {\small Princeton University, Princeton, NJ 08540, USA}\\
 {\small $^*$ Corresponding author email address: jessejenkins@princeton.edu}
}

\date{}

\maketitle

\section{Building 6-Zone Capacity Expansion Model with {\tt Plasmo.jl}}

Here we give an overview of modeling the 6-zone capacity expansion model (CEM) using the RemoteOptiGraph abstraction in {\tt Plasmo.jl} \cite{jalving2022graph}. We will show a comparison of code using a regular (i.e., non-distributed) {\tt OptiGraph} and then show code using a {\tt RemoteOptiGraph}. These sets of code rely on source code from {\tt GenX.jl} \cite{GenX} for constructing models and which is adapted from \cite{cole2025graph}. We will also discuss some nuances in writing performant code with {\tt RemoteOptiGraph}s

We also note that there are several ways that the code for building these problems can be structured. Our choice of defining functions that construct different parts of the code is one of many options and was chosen because of how the source code of {\tt GenX.jl} is structured. The functions that we discuss below are implemented in the source code of {\tt GenX.jl} (not an official registered version) which has been adapted for working with {\tt Plasmo.jl} and is provided in the repository \cite{Zenodo_data}. However, the principles contained here are applicable to building other models in {\tt Plasmo.jl}. The variables, constraints, and objectives are largely added to the {\tt OptiGraph}s or {\tt RemoteOptiGraph}s through functions within {\tt GenX.jl}, so we focus on the high-level details and principles of how the {\tt OptiGraph}s or {\tt RemoteOptiGraph}s can be constructed. 

\subsection{Building the OptiGraph}

To build the 6-zone CEM using an {\tt OptiGraph} in {\tt Plasmo.jl}, we introduce functions in Code Snippets \ref{code:optigraph_snippet1} and \ref{code:optigraph_snippet2} which use functions from {\tt GenX.jl} to create the optimization problem. The construction of the OptiGraph is done in the function {\tt generate\_optigraph\_model} (Code Snippet \ref{code:optigraph_snippet1} which takes inputs of dictionaries {\tt setup}, {\tt inputs}, and {\tt inputs\_decomp} as well as optimizers and other meta-data). These dictionaries are specific to {\tt GenX.jl} and are information loaded from .csv and .yml files. These files contain problem data and can be found in the public repository \cite{Zenodo_data}. The general approach of this function is to build each subproblem as a graph independently, add these subgraphs to an overall graph, and then to add linking constraints between the subgraphs. Within this function, {\tt OptiGraph}s are instantiated for the overall graph ({\tt graph} in Line \ref{line:s1_graph}), the planning subgraph ({\tt planning\_graph} in Line \ref{line:s1_pgraph}) and the operation subgraphs ({\tt operation\_graphs} in Line \ref{line:s1_ographs}). A node is added to {\tt planning\_graph} (Line \ref{line:s1_pnode}) and to each of the {\tt operation\_graphs} (Lines \ref{line:s1_onodes1} - \ref{line:s1_onodes2}). The function {\tt load\_planning\_optinode!} is then called to add the planning model (Line \ref{line:s1_load_pnode}). A mapping of subproblems to their planning variables is also created for facilitating the linking constraints between the planning level and operations level (Line \ref{line:s1_var_map}). Next, the operation models are added to their respective nodes (Lines \ref{line:s1_load_onode1} - \ref{line:s1_load_onode2}) by calling {\tt load\_operation\_optinode!}. Importantly, this is parallelized by calling {\tt Threads.@threads} at the start of the loop since each graph can be constructed independently of the others. After this is complete, each of the subgraphs are added to the overall graph and optimizers are set (Lines \ref{line:s1_addgraphs1} - \ref{line:s1_addgraphs2}) and linking constraints are added via the function {\tt make\_optigraph\_linking\_constraints!} (Line \ref{line:s1_add_links}).

\begin{figure}[!htp]
    \begin{minipage}[t]{0.9\linewidth}
        \begin{scriptsize}
        \lstset{language=Julia, breaklines = true}
        \begin{lstlisting}[label = code:optigraph_snippet1, caption = Code containing the function for building the OptiGraph of the GenX.jl capacity expansion model.] 
function generate_optigraph_model(setup::Dict,inputs::Dict,PLANNING_OPTIMIZER::MOI.OptimizerWithAttributes,SUBPROB_OPTIMIZER::MOI.OptimizerWithAttributes, num_subproblems, inputs_decomp; remote::Bool = false, worker_list = [])

    # If `remote` == true, then build the RemoteOptiGraph; otherwise, use normal OptiGraphs
    if !(remote)
        graph = OptiGraph(); |\label{line:s1_graph}|
        planning_graph = OptiGraph() |\label{line:s1_pgraph}|
        operation_graphs = [OptiGraph() for i in 1:num_subproblems] |\label{line:s1_ographs}|
    else
        graph = RemoteOptiGraph(worker = 1) |\label{line:s1_remotegraph1}|
        planning_graph = RemoteOptiGraph(worker = 1)
        operation_graphs = [RemoteOptiGraph(worker = worker_list[i]) for i in 1:num_subproblems] |\label{line:s1_remotegraph2}|
    end

    # Define a node on the `planning_graph`
    @optinode(planning_graph, planning_node) |\label{line:s1_pnode}|

    # Define a set of nodes on each `operation_graph`
    operation_nodes = [] |\label{line:s1_onodes1}|
    for i in 1:num_subproblems
        node = @optinode(operation_graphs[i], operation_node)
        push!(operation_nodes, node)
    end |\label{line:s1_onodes2}|

    # Load the planning node data from GenX
    planning_variables = GenX.load_planning_optinode!(planning_node,setup,inputs) |\label{line:s1_load_pnode}|

    # Define a map of variables from the planning graph to subgraphs
    linking_variables_maps = Dict{Int64,Dict{AbstractVariableRef, AbstractVariableRef}}() |\label{line:s1_var_map}|

    # build the operations subgraphs asynchronously
    Threads.@threads for i in 1:num_subproblems |\label{line:s1_load_onode1}|
        linking_variables_maps[i] = GenX.load_operation_optinode_remote!(operation_graphs[i],setup,inputs_decomp[i],planning_variables)
    end |\label{line:s1_load_onode2}|

    # Set optimizer and add all the graphs to the overall graph
    set_optimizer(planning_graph, PLANNING_OPTIMIZER) |\label{line:s1_addgraphs1}|
    add_subgraph(graph, planning_graph)
    set_to_node_objectives(planning_graph)
    for i in 1:num_subproblems
        set_optimizer(operation_graphs[i], SUBPROB_OPTIMIZER)
        add_subgraph(graph, operation_graphs[i])
        set_to_node_objectives(operation_graphs[i])
    end |\label{line:s1_addgraphs2}|

    # Define linking constraints between subgraphs
    GenX.make_optigraph_linking_constraints!(graph,linking_variables_maps) |\label{line:s1_add_links}|

    return graph,linking_variables_maps
end
        \end{lstlisting}
        \end{scriptsize}
    \end{minipage}
\end{figure}

The functions {\tt load\_planning\_optinode!} (Line \ref{line:s2_load_pnode}), {\tt load\_operation\_optinode!} (Line \ref{line:s2_load_onode}), and {\tt make\_optigraph\_linking\_constraints!} (Line \ref{line:s2_linking}) are shown in Code Snippet \ref{code:optigraph_snippet2}. The first function calls {\tt GenX.jl}'s internal function {\tt planning\_model!} (Line \ref{line:s2_planning_model}). This function adds all of the variables and constraints for representing the planning level of the 6-zone CEM. The function {\tt load\_planning\_optinode!} also adds a variable and expression that are used by this function internally (Lines \ref{line:s2_pnode_var1} - \ref{line:s2_pnode_var2}). Similarly, the function {\tt load\_operation\_optinode!} calls {\tt GenX.jl}'s internal function {\tt operation\_model!} (Line \ref{line:s2_operation_model}) which adds the variables and constraints for representing the operations subproblems. The function {\tt load\_operation\_optinode!} also contains functionality for mapping the names of the varibles to the planning level variables (Lines \ref{line:s2_mapping1} - \ref{line:s2_mapping2}). This mapping is returned by the function and used by the function {\tt make\_optigraph\_linking\_constraints!} to add linking constraints (Line \ref{line:s2_linking_cons}) to the graph between the planning level and operational subproblems.

\begin{figure}[!htp]
    \begin{minipage}[t]{0.9\linewidth}
        \begin{scriptsize}
        \lstset{language=Julia, breaklines = true}
        \begin{lstlisting}[label = code:optigraph_snippet2, caption = Code containing the function for building the planning and operation subproblems of the OptiGraph.] 
function load_planning_optinode!(EP::N,setup::Dict,inputs::Dict) where {N <: Union{OptiNode, RemoteNodeRef}} |\label{line:s2_load_pnode}|
    # Define some variables and expressions on the planning node
    num_subperiods = inputs["REP_PERIOD"];
    @variable(EP, vZERO==0) |\label{line:s2_pnode_var1}|
    V = typeof(vZERO)
    @expression(EP,eObj,GenericAffExpr{Float64, V}(0.0))|\label{line:s2_pnode_var2}|

    # Call the GenX function to build the planning model
    GenX.planning_model!(EP,setup,inputs) |\label{line:s2_planning_model}|

    # Update the objective
    @variable(EP,vTHETA[1:num_subperiods]>=0)
    @objective(EP, Min, setup["ObjScale"]*(EP[:eObj]+sum(EP[:vTHETA])))
    planning_vars = setdiff(all_variables(EP),[EP[:vZERO];EP[:vTHETA]]);
    return planning_vars
end

function load_operation_optinode!(EP::N,setup::Dict,inputs::Dict,planning_variables::Vector{T}; remote::Bool=false) where {N <: Union{OptiNode, RemoteNodeRef}, T <: AbstractVariableRef} |\label{line:s2_load_onode}|

    # Define some variables and expressions on the planning node
    @variable(EP, vZERO==0)
    V = typeof(vZERO)
    @expression(EP,eObj,GenericAffExpr{Float64, V}(0.0))

    # Call the GenX function to build the operation model
    GenX.operation_model!(EP,setup,inputs) |\label{line:s2_operation_model}|

    #Update objective
    @objective(EP, Min, setup["ObjScale"]*EP[:eObj])

    # Build the set of linking variables; remoteoptigraphs have slightly different names under the current version
    all_vars = all_variables(EP); |\label{line:s2_mapping1}|
    var_strings = name.(all_vars)
    if !remote
        clean_var_strings = [replace(str,r"operation_node\[:(.*)\]"=>s"\1") for str in var_strings]
        planning_var_strings = name.(planning_variables)
        clean_planning_var_strings = [replace(str,r"planning_node\[:(.*)\]"=>s"\1") for str in planning_var_strings]
    else
        clean_var_strings = [replace(str,r"operation_node\[operation_node" * r"\[:(.*)\]\]"=>s"\1") for str in var_strings] |\label{line:s2_remote_parse1}|
        planning_var_strings = name.(planning_variables)
        clean_planning_var_strings = [replace(str,r"planning_node\[planning_node\[:(.*)\]\]"=>s"\1") for str in planning_var_strings] |\label{line:s2_remote_parse2}|
    end

    common_subproblem_indices = [findfirst(clean_var_strings.==s) for s in intersect(clean_var_strings,clean_planning_var_strings)]
    linking_variables_map = Dict(all_vars[i]=>planning_variables[findfirst(clean_planning_var_strings.==clean_var_strings[i])] for i in common_subproblem_indices) |\label{line:s2_mapping2}|

    return linking_variables_map
end

function make_optigraph_linking_constraints!(graph::Plasmo.AbstractOptiGraph,linking_variables_maps::Dict{Int64,Dict{V,V}}) where{V <: JuMP.AbstractVariableRef} |\label{line:s2_linking}|
    # Add linking constraints between planning and operations subgraphs
    for i in keys(linking_variables_maps)
        for k in keys(linking_variables_maps[i])
            @linkconstraint(graph, k == linking_variables_maps[i][k]) |\label{line:s2_linking_cons}|
        end
    end
end
        \end{lstlisting}
        \end{scriptsize}
    \end{minipage}
\end{figure}

\subsection{Building the RemoteOptiGraph}

In this section, we outline two different ways that the RemoteOptiGraph can be constructed. The first method is to use virtually the same code as used for a regular {\tt OptiGraph} but instead instantiating the graph with the constructor {\tt RemoteOptiGraph} instead of {\tt OptiGraph}. This is shown in Code Snippets \ref{code:optigraph_snippet1} and \ref{code:optigraph_snippet2}, where the {\tt remote} flag is set to {\tt true} when passed to the {\tt generate\_optigraph\_model} function in Code Snippet \ref{code:optigraph_snippet1}. When this is done, the function will instantiate using the {\tt RemoteOptiGraph} (Lines \ref{line:s1_remotegraph1} - \ref{line:s1_remotegraph2}). The remainder of the code follows as before (with the exception of the parsing of variable names; see Lines \ref{line:s2_remote_parse1} - \ref{line:s2_remote_parse2} in Code Snippet \ref{code:optigraph_snippet2}) because the {\tt JuMP.jl} functions like {\tt @expression}, {\tt @variable}, {\tt @constraint}, or {\tt @objective} are extended for the {\tt RemoteOptiGraph}. The source code for {\tt planning\_model!} and {\tt operation\_model!} in {\tt GenX.jl} similarly use these {\tt JuMP.jl} functionalities so that the source code undergoes minimal change for using the {\tt OptiGraph} or the {\tt RemoteOptiGraph}. 

One challenge with using the {\tt RemoteOptiGraph} using the same code as the {\tt OptiGraph} is that it can be slow. Each call of the {\tt JuMP.jl} macros typically includes at least one {\tt fetch} call to the remote worker. Each of these calls requires serialization and deserialization of information between workers, creating significant overhead. The source code of {\tt GenX.jl} includes many such calls, and is then much slower to build the {\tt RemoteOptiGraph} in this manner. 

A second approach for constructing the {\tt RemoteOptiGraph} which can be much faster is to call the {\tt planning\_model!} and {\tt operation\_model!} functions of {\tt GenX.jl} directly on the remote workers (Code Snippets \ref{code:remoteoptigraph_snippet3} - \ref{code:remoteoptigraph_snippet5}), avoiding much of the communication overhead. This is facilitated by importing {\tt GenX.jl} directly on each worker by calling {\tt @everywhere using GenX} once all the remote workers have been added (see scripts for running the models at \cite{Zenodo_data}). Once this is done, the function {\tt generate\_remote\_optigraph\_model} (Code Snippet \ref{code:remoteoptigraph_snippet3}) can be called on the main worker to create the RemoteOptiGraph. This function has a similar form to those introduced in Code Snippets \ref{code:optigraph_snippet1} and \ref{code:optigraph_snippet2}, and we will next walk through the differences between these functions. 

The {\tt RemoteOptiGraph} is created in Code Snippet \ref{code:remoteoptigraph_snippet3} in the function {\tt generate\_remote\_optigraph\_model}, which has a similar form to {\tt generate\_optigraph\_model}. Within this function, {\tt RemoteOptiGraph}s are instantiated (Lines \ref{line:s3_graph1} - \ref{line:s3_graph2}), the planning model is loaded via the {\tt load\_planning\_optinode\_remote!} call (Line \ref{line:s3_pnode}), and a linking variable map is created (Line \ref{line:s3_var_map}). Because each subgraph may be on a different worker, we can form each operation subgraph asynchronously by placing the {\tt load\_operation\_optinode\_remote!} call inside an {\tt @async} call (Lines \ref{line:s3_async1} - \ref{line:s3_async2}) in the for loop. This allows the for loop to continue to the next iteration without the code inside the {\tt @async} call completing. This for loop is also wrapped within an {\tt @sync} call (Line \ref{line:s3_sync}) which ensures the code run by the for loop completes before proceeding to the rest of the function. After this, the subgraphs are added to the overall graph (Lines \ref{line:s3_add_subgraph1} - \ref{line:s3_add_subgraph2}) and the linking constraints are added (Line \ref{line:s3_linking_cons}).

\begin{figure}[!htp]
    \begin{minipage}[t]{0.9\linewidth}
        \begin{scriptsize}
        \lstset{language=Julia, breaklines = true}
        \begin{lstlisting}[label = code:remoteoptigraph_snippet3, caption = Code containing the function for building the RemoteOptiGraph of the GenX.jl capacity expansion model.] 
function generate_remote_optigraph_model(setup::Dict,inputs::Dict,PLANNING_OPTIMIZER::MOI.OptimizerWithAttributes,SUBPROB_OPTIMIZER::MOI.OptimizerWithAttributes, num_subproblems, inputs_decomp; worker_list = [])
    # Define RemoteOptiGraph
    graph = RemoteOptiGraph(); |\label{line:s3_graph1}|
    planning_graph = RemoteOptiGraph()
    operation_graphs = [RemoteOptiGraph(worker = worker_list[i]) for i in 1:num_subproblems] |\label{line:s3_graph2}|

    # Build planning node
    planning_variables = GenX.load_planning_optinode_remote!(planning_graph,setup,inputs) |\label{line:s3_pnode}|

    # Define a map of variables from the planning graph to operations graphs
    linking_variables_maps = Dict{Int64,Dict{Plasmo.RemoteVariableRef,Plasmo.RemoteVariableRef}}() |\label{line:s3_var_map}|

    # Build the operations subgraphs asynchronously
    @sync begin |\label{line:s3_sync}|
        for i in 1:num_subproblems
            @async begin|\label{line:s3_async1}|
                linking_variables_maps[i] = GenX.load_operation_optinode_remote!(operation_graphs[i],setup,inputs_decomp[i],planning_variables)
            end|\label{line:s3_async2}|
        end
    end 

    # Set optimizer and add all the graphs to the overall graph
    set_optimizer(planning_graph, PLANNING_OPTIMIZER) |\label{line:s3_add_subgraph1}|
    add_subgraph(graph, planning_graph)
    for i in 1:num_subproblems
        Plasmo.add_subgraph(graph, operation_graphs[i])
        JuMP.set_optimizer(operation_graphs[i], SUBPROB_OPTIMIZER)
    end |\label{line:s3_add_subgraph2}|

    # Define linking constraints between subgraphs
    GenX.make_optigraph_linking_constraints!(graph,linking_variables_maps) |\label{line:s3_linking_cons}|

    return graph,linking_variables_maps
end
        \end{lstlisting}
        \end{scriptsize}
    \end{minipage}
\end{figure}

The {\tt load\_planning\_optinode\_remote!} function is outlined in Code Snippet \ref{code:remoteoptigraph_snippet4}. The bulk of this function is run in a remote call in the {\tt @spawnat} macro. To begin, the {\tt DArray} (distributed array from {\tt DistributedArrays.jl} \cite{distributedarrays}) containing the {\tt OptiGraph} on the remote worker is retrieved from the {\tt RemoteOptiGraph} object (Line \ref{line:s4_darray}). The {\tt @spawnat} call begins by setting the worker on which the call will run, which worker is stored in the {\tt worker} field of the {\tt RemoteOptiGraph} (Line \ref{line:s4_spawnat_begin}). Code contained inside the {\tt @spawnat} macro will run on that worker (Lines  \ref{line:s4_spawnat_begin} - \ref{line:s4_spawant_end}). Inside this code block, the {\tt OptiGraph} is retrieved from the {\tt DArray} (Line \ref{line:s4_lgraph}) via the function {\tt local\_graph}. The node which will contain the planning model is then directly added to this graph (Line \ref{line:s4_pmodel}), and, as with the {\tt load\_planning\_optinode!} function, variables, expressions, and objectives are defined and the {\tt planning\_model!} function is called (Lines \ref{line:s4_pbuild1} - \ref{line:s4_pbuild2}). While the {\tt load\_planning\_optinode!} function returns a set of variables, the variables of interest in the {\tt RemoteOptiGraph} case are on the remote worker and must be fetched. This is done by getting the variables of interest on the remote worker (Line \ref{line:s4_lvars}) and converting them to {\tt ProxyVarRef}s through {\tt Plasmo.jl}'s function {\tt \_convert\_local\_to\_proxy} (Line \ref{line:s4_pvars}), which are then fetched to the main worker (Line \ref{line:s4_fetch}) and converted to {\tt RemoteVariabelRef}s (Line \ref{line:s4_rvars}) and returned. 

\begin{figure}[!htp]
    \begin{minipage}[t]{0.9\linewidth}
        \begin{scriptsize}
        \lstset{language=Julia, breaklines = true}
        \begin{lstlisting}[label = code:remoteoptigraph_snippet4, caption = Code containing the function for building the planning RemoteOptiGraph of the GenX.jl capacity expanion model] 
function load_planning_optinode_remote!(rgraph::Plasmo.RemoteOptiGraph,setup::Dict,inputs::Dict)
    num_subperiods = inputs["REP_PERIOD"];
    # Get the DistributedArray of the RemoteOptiGraph; this is a pointer to the OptiGraph
    darray = rgraph.graph |\label{line:s4_darray}|

    # Run the code on the remote worker
    f = @spawnat rgraph.worker begin |\label{line:s4_spawnat_begin}|
        # Get the OptiGraph
        lgraph = Plasmo.local_graph(darray) |\label{line:s4_lgraph}|

        # Define the node and additional variables
        @optinode(lgraph, planning_node) |\label{line:s4_pnode}|
        @variable(planning_node, vZERO==0) |\label{line:s4_pbuild1}|
        @expression(planning_node,eObj,GenericAffExpr{Float64, Plasmo.NodeVariableRef}(0.0))

        # Call the GenX function to build the planning model
        GenX.planning_model!(planning_node,setup,inputs) |\label{line:s4_pmodel}|

        # Update the objective
        @variable(planning_node,vTHETA[1:num_subperiods]>=0)
        @objective(planning_node, Min, setup["ObjScale"]*(planning_node[:eObj]+sum(planning_node[:vTHETA])))
        set_to_node_objectives(lgraph) |\label{line:s4_pbuild2}|

        # Return the information needed to create the RemoteVariableRefs
        planning_vars = setdiff(all_variables(planning_node),[planning_node[:vZERO];planning_node[:vTHETA]]); |\label{line:s4_lvars}|
        Plasmo._convert_local_to_proxy(lgraph, planning_vars) |\label{line:s4_pvars}|
    end |\label{line:s4_spawant_end}|
    # Fetch proxy variables and convert to remote
    pvars = fetch(f) |\label{line:s4_fetch}|
    rvars = Plasmo._convert_proxy_to_remote(rgraph, pvars) |\label{line:s4_rvars}|

    return rvars
end
        \end{lstlisting}
        \end{scriptsize}
    \end{minipage}
\end{figure}

The {\tt load\_operation\_optinode\_remote!} in Code Snippet \ref{code:remoteoptigraph_snippet5} has a similar format. The {\tt DArray} is likewise retrieved from the {\tt RemoteOptiGraph} (Line \ref{line:s5_darray}), and the bulk of the code is run on the remote worker inside an {\tt @spawnat} call (Lines \ref{line:s5_spawnat_begin} - \ref{line:s5_spawnat_end}). The {\tt OptiGraph} is retrieved from the {\tt DArray} (Line \ref{line:s5_lgraph}), and the node containing the operation model is instantiated directly on the {\tt OptiGraph} (Line \ref{line:s5_onode}). Variables, expressions, and objectives are added to the node and the {\tt operation\_model!} function is called (Lines \ref{line:s5_obuild1} - \ref{line:s5_obuild2}). As with the {\tt load\_operation\_optinode!} function, this function parses variable names and creates a set of variables and returns a dictionary. However, this parsing is done on the remote worker (Lines \ref{line:s5_parse1} - \ref{line:s5_parse2}), and a set of planning indices is also fetched from the remote worker along with the {\tt ProxyVarRef}s (Lines \ref{line:s5_fetch1} and \ref{line:s5_fetch2}) which is used to form the final dictionary (Line \ref{line:s5_final_dict}).

\begin{figure}[!htp]
    \begin{minipage}[t]{0.9\linewidth}
        \begin{scriptsize}
        \lstset{language=Julia, breaklines = true}
        \begin{lstlisting}[label = code:remoteoptigraph_snippet5, caption = Code containing the function for building the operations RemoteOptiGraph of the GenX.jl capacity expanion model] 
function load_operation_optinode_remote!(rgraph::RemoteOptiGraph,setup::Dict,inputs::Dict,planning_variables::Vector{Plasmo.RemoteVariableRef})
    # Get the DistributedArray of the RemoteOptiGraph; this is a pointer to the OptiGraph
    darray = rgraph.graph |\label{line:s5_darray}|

    planning_var_strings = JuMP.name.(planning_variables)
    # Run the code on the remote worker
    f = @spawnat rgraph.worker begin |\label{line:s5_spawnat_begin}|
        lgraph = Plasmo.local_graph(darray) |\label{line:s5_lgraph}|

        # Define the node and additional variables
        Plasmo.@optinode(lgraph, operation_node) |\label{line:s5_onode}|
        JuMP.@variable(operation_node, vZERO==0) |\label{line:s5_obuild1}|
        JuMP.@expression(operation_node,eObj,JuMP.GenericAffExpr{Float64, Plasmo.NodeVariableRef}(0.0))

        # Call the GenX function to build the planning model
        GenX.operation_model!(operation_node,setup,inputs)

        # Update the objective
        JuMP.@objective(operation_node, Min, setup["ObjScale"]*operation_node[:eObj])
        Plasmo.set_to_node_objectives(lgraph) |\label{line:s5_obuild2}|

        # Build the set of linking variables by matching string names
        all_vars = JuMP.all_variables(operation_node); |\label{line:s5_parse1}|
        var_strings = JuMP.name.(all_vars)

        clean_var_strings = [replace(str,r"operation_node\[:(.*)\]"=>s"\1") for str in var_strings]
        #planning_var_strings = name.(planning_variables)
        clean_planning_var_strings = [replace(str,r"planning_node\[planning_node\[:(.*)\]\]"=>s"\1") for str in planning_var_strings]

        common_variable_names = intersect(clean_var_strings,clean_planning_var_strings)
        common_subproblem_indices = [findfirst(clean_var_strings.==s) for s in common_variable_names] #return indicies of clean_var_strings that match a planning variable
        common_planning_indices = [findfirst(clean_planning_var_strings.==s) for s in common_variable_names] #return indicies of clean_var_strings that match a planning variable |\label{line:s5_parse2}|

        # Convert local variables to proxy for fetching
        lvars = [all_vars[i] for i in common_subproblem_indices]
        pvars = Plasmo._convert_local_to_proxy(lgraph, lvars)

        pvars, common_planning_indices |\label{line:s5_fetch1}|
    end |\label{line:s5_spawnat_end}|
    # Fetch proxy variables and indices of planning names and return mapping
    pvars, cpi = fetch(f) |\label{line:s5_fetch2}|
    rvars = Plasmo._convert_proxy_to_remote(rgraph, pvars)
    
    return Dict(rvars[i] => planning_variables[idx] for (i, idx) in enumerate(cpi)) |\label{line:s5_final_dict}|
end
        \end{lstlisting}
        \end{scriptsize}
    \end{minipage}
\end{figure}

This second approach to building the {\tt RemoteOptiGraph} is slightly more complex, but, for this problem is significantly faster in building the optimization problem. The primary change in this second approach from the first is that the calls to {\tt planning\_model!} and {\tt operation\_model!} from {\tt GenX.jl} are done inside an {\tt @spawnat} call, allowing each subgraph to be constructed asynchronously. For other use cases, users can define such functions globally for nodes or optigraphs by defining such functions inside {\tt @everywhere} calls. Alternatively, users can choose to avoid writing their own remote calls by using the {\tt JuMP.jl} macros that have been extended for the {\tt RemoteOptiGraph}s at the cost of higher model construction overhead. 

\section{Running on Distributed Architectures}

The functions outlined above operate independently of the computing architecture used because {\tt Plasmo.jl} uses {\tt Distributed.jl} \cite{bezanson2017julia} which is robust for working under different distributed computing architectures. In this work, we ran the 6-zone CEM on three different architectures: i) a single node on a computing cluster with multiple processors; ii) multiple nodes on a Slurm \cite{slurm} computing cluster with multiple processors; iii) a cloud computing architecture. The source code for building the model stayed the same across the three different architectures, and the primary difference for running on the three architectures was that they used different methods for adding remote processes to the main worker.

All three architectures use {\tt Distributed.jl}'s {\tt addprocs} function to specify the distributed computing environment. For the first architecture, local processes are added directly using the default {\tt
addprocs} function. For the second architecture, Slurm processes are added via {\tt SlurmManager} from the {\tt ClusterManagers.jl} package, which interfaces with the HPC-specific job scheduling system. 
 Finally we used the cloud computing architecture to demonstrate how our distributed architecture allows for enable elastic scaling and rapid deployment of workers outside the HPC environment. The workers were provisioned using {\tt Terraform}, configured with {\tt Ansible}, and the necessary Julia environment containerized with {\tt Docker}, after which their IP addresses were passed to {\tt addprocs} to connect to the remote workers via SSH. Further details on this implementation are provided in the repository’s documentation and examples \cite{Zenodo_data}.


\bibliography{DistributedPlasmo}